\documentclass[twocolumn]{aastex63_ionthinspace}

\usepackage{xcolor}
\usepackage{amsmath}
\usepackage{affils}

\newcommand{\camath}{C/A}

\newcommand{\HI}{\ion{H}{1}}

\newcommand{\HH}{${\rm H_2}$}
\newcommand{\hi}[0]{\HI{}}

\newcommand{\reff}[0]{${\rm R_{\rm eff}}$}

\newcommand{\thh}[0]{$^{\rm th}$}
\newcommand{\kms}[0]{{\ \rm km\ s}^{-1}}
\newcommand{\galex}[0]{\textit{GALEX}}

\newcommand{\rrr}[1]{\textcolor{black}{#1}}
\newcommand{\rrrtwo}[1]{\textcolor{black}{#1}}

\newcommand{\Ytot}[0]{$\Upsilon_{\rm tot}$}
\newcommand{\citeok}[0]{\cite{ostriker2022}} 
\newcommand{\citealtok}[0]{\citealt{ostriker2022}}

\newcommand{\oten}[0]{PRFM}
\newcommand{\alphaco}[0]{$\alpha_{\rm CO}$}
\newcommand{\pde}[0]{$P_{\rm DE}$}
\newcommand{\sigsfr}[0]{$\Sigma_{\rm SFR}$}
\defcitealias{paperone}{Paper I}
\begin{document}
\shortauthors{Kado-Fong et al.}

\title{Ultra-Diffuse Galaxies as Extreme Star-forming Environments II:\\ Star Formation and Pressure Balance in \HI{}-Rich UDGs}

\DeclareAffil{princeton}{Department of Astrophysical Sciences, Princeton University,Princeton, NJ 08544, USA}

\author[0000-0002-0332-177X]{Erin Kado-Fong}
\affiliation{Department of Astrophysical Sciences, Princeton University,Princeton, NJ 08544, USA}
\author[0000-0003-2896-3725]{Chang-Goo Kim}
\affiliation{Department of Astrophysical Sciences, Princeton University,Princeton, NJ 08544, USA}
\author[0000-0002-5612-3427]{Jenny E. Greene}
\affiliation{Department of Astrophysical Sciences, Princeton University,Princeton, NJ 08544, USA}
\author[0000-0002-0041-4356]{Lachlan Lancaster}
\affiliation{Department of Astrophysical Sciences, Princeton University,Princeton, NJ 08544, USA}
\correspondingauthor{Erin Kado-Fong} 
\email{kadofong@princeton.edu}
  
  \date{\today}

\begin{abstract}
In addition to occupying the extreme, diffuse tail
of the dwarf galaxy population, 
Ultra-Diffuse Galaxies (UDGs) are
themselves a key laboratory in which to study star formation in extreme low-density environments.
In the second paper of this series, we compare the  
spatially resolved star formation activity of 
22 \HI-selected UDGs and 21 ``normal'' dwarf galaxies within 120 Mpc 
to predictions within
the pressure-regulated, feedback-modulated (PRFM) theory of star formation. 
To do so, we employ
a joint SED fitting method that allows us to estimate star formation rate
and stellar mass surface density from UV-optical imaging. 
We find that the PRFM framework extends successfully to the UDG regime -- 
although the UDGs in our sample show unusually low star
formation rate surface densities given their \HI{} content, this low
star formation efficiency can be naturally explained by the diffuse structure of
the UDGs. In fact, when cast in the PRFM framework, 
the relationship between midplane pressure and star formation in the UDG sample
is in good agreement not only with the ``normal'' dwarf reference sample, but also
with measurements from more massive galaxies. Our results suggest that 
despite their low star formation efficiencies, the
\HI-rich UDGs need not be forming stars in an exotic manner.
We also find that 
the UDGs are likely \HH-poor compared even to the overall dwarf population.
\end{abstract}

\section{Introduction}
Star formation and galaxy evolution are 
intrinsically linked processes; a full understanding of one is not possible in the 
absence of a theory of the other. In addition to the most definitional link -- that
star formation grows stellar mass while consuming gas -- the resulting feedback from
star formation also has a direct impact on the structure of the galaxy's interstellar
medium (ISM) via processes including
supernovae, UV radiation, and
stellar winds \citep[see, e.g.][]{kim2013, kim2017, girichidis2018, kannan2019, kim2021, lancaster2021}. 
In order to understand how and why galaxies evolve the way that they do
-- especially at the low-mass end, where star formation feedback is expected to
play an outsized role \citep[see, e.g.][]{silk1997, dekel1986, elbadry2016, behroozi2019, hu2019, dashyan2020, smith2020} -- we must understand the self-regulatory process of star formation. 
Likewise, to understand the environments in which star formation proceeds, we must understand the arc of galaxy evolution through cosmic time.

\begin{figure*}[htb]
\centering     
\includegraphics[width=\linewidth]{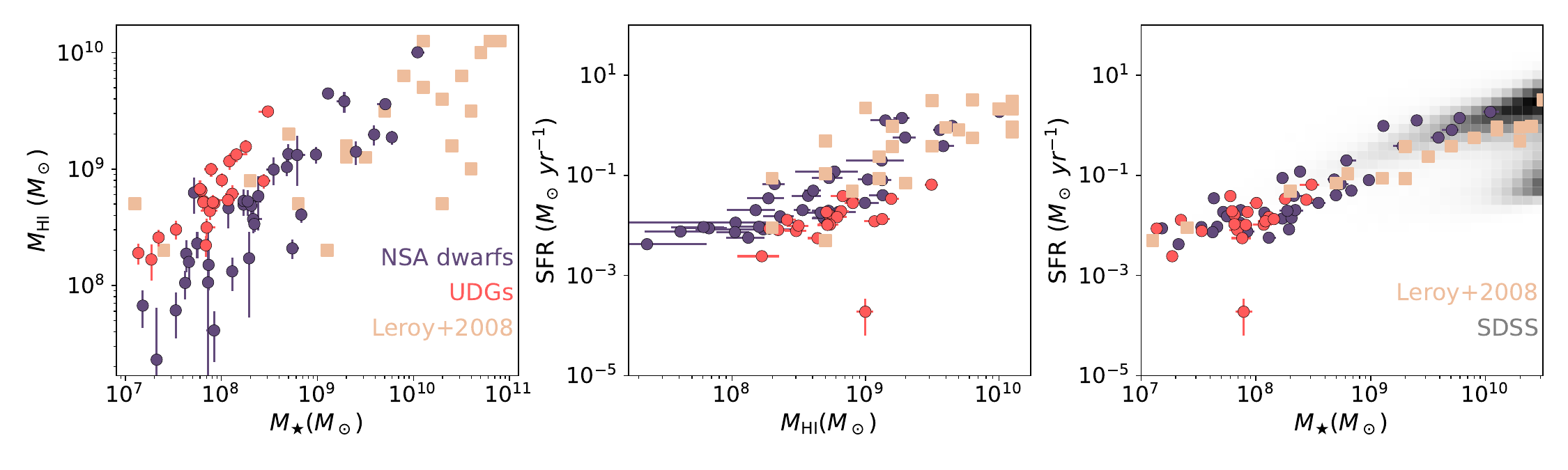}
\caption{ 
    A comparison of the integrated properties of the samples derived in this work
    against the directly measured result of \cite{leroy2008} (beige points). In all
    panels, the red points show UDGs and the purple points show NASA Sloan Atlas (NSA, our ``normal'' dwarf comparison
    sample -- see text) dwarfs. From left,
    we show the relationship between stellar mass and \HI{} mass, the relationship between
    \HI{} mass and SFR, and the star-forming main sequence (SFMS). In the SFMS panel, 
    we also show the results of the SDSS DR7 MPA-JHU added-value catalogs in greyscale
    \citep{kauffmann2003, brinchmann2004, salim2007}. We find a good agreement between
    our results and those from the literature. Although the UDGs have high \HI{} masses
    for their stellar mass (left panel), they have relatively little SFR for their \HI{}
    mass (middle panel).
    }
\label{f:gascomparison}
\end{figure*}

In both the fields of star formation and galaxy evolution, we seek out ``extreme''
cases in order to best test and stretch 
our understanding of the underlying physical processes at work.
Ultra-diffuse galaxies (UDGs)
are dwarf galaxies characterized by large stellar sizes 
and low surface brightnesses. 
Though exact definitions vary, UDGs are typically required
to have effective radii greater than 1.5 kpc and 
central surface brightnesses fainter than 24 mag arcsec$^{-2}$
(see \citealt{vannest2022} for an overview of UDG definitions).
These diffuse galaxies are extreme as both a
product of galaxy evolution and as an environment 
in which stars form.
Indeed, the study of UDGs as an extreme sector of the dwarf galaxy population has
enjoyed both extensive observational
\citep[see, e.g.][]{sandage1984, mcgaugh1995, dalcanton1997, vandokkum2015, beasley2016, beasley2016b, peng2016, yagi2016, leisman2017, greco2018a, greco2018b, vandokkum2018, danieli2019, janowiecki2019, vandokkum2019, danieli2021, gault2021, greene2022} and theoretical \citep[e.g.][]{amorisco2016, dicintio2017, chan2018, jiang2019, liao2019, wright2021, vannest2022} study. 
However, less work has been devoted to their star formation properties.

UDGs are extreme environments for star formation due to their presumably low
stellar mass surface densities and likely shallow potential wells (for UDGs in the field,
see, for example, \cite{leisman2017} and \cite{kong2022}
-- typical halo masses for UDGs in clusters are a topic of
significant debate, see, e.g., \citealt{vandokkum2015, beasley2016, vandokkum2019, sales2020, saifollahi2021}). These conditions are a
marked departure from Solar Neighborhood-like or even outer (Milky Way) disk-like 
conditions that serve as the fiducial environment for many models of star formation 
\citep{krumholz2009, ostriker2010}. 
Mapping out the star formation activity in UDGs 
thus serves as a test of the extensibility of these star formation models to new and
extreme diffuse conditions.

In the first paper of this series 
\citep[][hereafter referred to as \citetalias{paperone}]{paperone} we demonstrated that
UDGs form stars inefficiently relative to their \HI{} surface densities (where 
SFE(\HI{})$\equiv \Sigma_{\rm SFR}/\Sigma_{\rm HI}$) on scales down to 500 pc.
However, a study of the atomic gas alone is insufficient to understand whether this 
low SFE(\HI) is unexpected given the diffuse structure of the UDGs. 
In \citetalias{paperone} we demonstrated that the star formation
in UDGs is different from normal dwarfs, but in this paper we will consider whether that
difference in star formation can be explained by our understanding of the physical
processes that drive and modulate star formation.

To this end, we turn to the pressure-regulated, feedback-modulated model of
star formation that has been developed over a series of works
\citep{ostriker2010, kim2011, ostriker2011, kim2013, kim2015} and most 
recently condensed into \citeok{}. 
This theory establishes a link between
star formation rate surface density 
and weight ($\mathcal{W}$) 
by considering the role of star formation feedback 
in maintaining the structure and energy density (pressure) of the ISM. This physical relationship provides us with
a method that links the two notable features of the UDGs in this work -- their
unusual structure and their low star formation rate surface density. With this framework
in hand, we will endeavour to explain the surprising -- or perhaps expected -- nature of
the star formation in \HI-rich UDGs.

We will structure the paper as follows, and note that readers familiar
with either \citetalias{paperone} or PRFM star formation may feel 
free to skip certain sections. In \autoref{s:datasets} we will 
summarize the data and methods that we utilized in \citetalias{paperone}
to produce the initial data products used in this analysis. 
We will then give an abbreviated overview of PRFM star formation 
theory in \autoref{s:prfm}, before moving onto a discussion of its
application to the dwarfs at hand in \autoref{s:results}. Finally,
we will discuss the implications of the analysis on our understanding
of the structure and star formation of UDGs in \autoref{s:discussion}

\begin{figure*}[htb]
\centering     
\includegraphics[width=\linewidth]{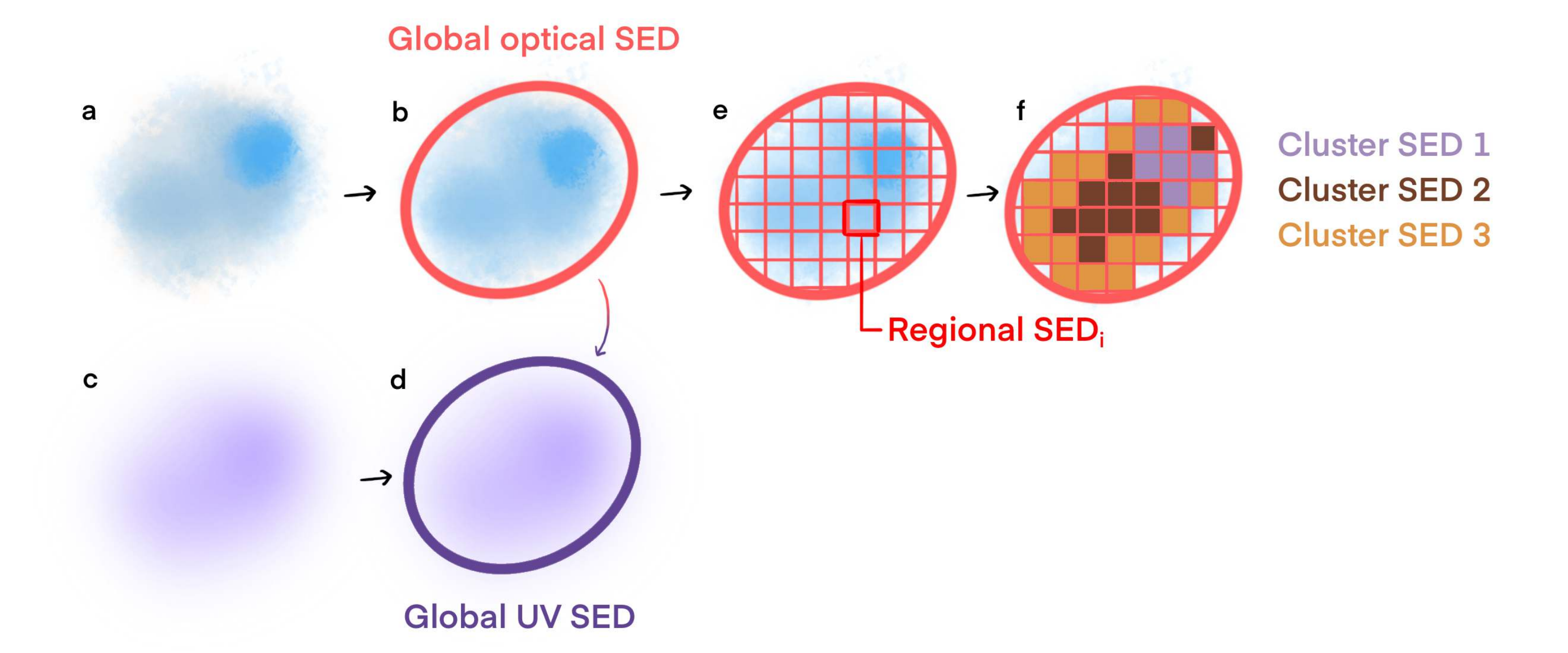}
\caption{ 
    A schematic layout of the fitting technique used in this work, as introduced in \citetalias{paperone}. 
    At left,
    we show a schematic representation of the optical data and
    UV data in panels \textit{a} and \textit{c}, respectively. In this
    analysis, the UV data are at much lower
    spatial resolution than the optical data. We then illustrate
    the global SED fit of the optical and 
    UV data in \textit{b} and \textit{d}; 
    the apertures here are held fixed over the UV and optical data.
    We then divide the optical data into 
    regions in which regional SEDs are measured (panel \textit{e}). 
    Panel \textit{f} shows the clustering of
    these spatial regions into three representative ``clusters''; these cluster SEDs will
    be fit jointly with the global UV photometry of panel \textit{d}.
    }
\label{f:schematic}
\end{figure*}
\section{Datasets}\label{s:datasets}
Readers familiar with \citetalias{paperone} of this series will find
that this content has been covered in greater detail in 
\citetalias{paperone}, and may skip to \autoref{s:prfm}. 
We provide the most salient
points of our sample here for those readers not familiar
with the first paper of this series, but encourage those 
readers with an interest in the methods to refer to 
\citetalias{paperone} for a more in-depth discussion of the 
analysis therein.

Our dwarf sample consists of two main branches: a high surface brightness, ``normal'' dwarf sample
drawn from the \HI{} catalog of \cite{bradford2015} -- hereafter the
NSA sample, and an ultra-diffuse galaxy sample 
with known distances from the \hi{} catalog of \cite{janowiecki2019}.
The NSA sample is drawn from the NASA Sloan-Atlas, a catalog 
of nearby galaxies re-analyzed from the Sloan Digital Sky 
Survey DR8 \citep{aihara2011, blanton2011}.
The UDG sample was selected to have a maximum distance of $d=120$ Mpc; we
enforce the same limit on the NSA sample. We note that in this work we will
refer to the ``normal'' galaxy sample as the NSA sample, as these galaxies were
not specifically chosen to exclude low surface brightness galaxies. Rather, their
relatively high surface brightnesses are a result of the observational selection which
lead to their inclusion in the NSA.

\subsection{Sample Overview}\label{s:methods:samples}

Let us first define the UDG sample. We draw our galaxies from the \HI-selected catalog
of \cite{janowiecki2019}, selecting those which have 5-band imaging from the
Hyper Suprime-Cam Subaru Strategic Program \citep[HSC-SSP,][]{aihara2019,aihara2022}.
Drawing from an \HI-selected sample, at least in the regime of the low surface brightness
UDGs, comes with two key advantages for the study of star formation in low-density conditions.
First, the galaxies selected via \HI{} observations tend to be relatively isolated, which is 
quite key given the large influence that environment -- and particularly the presence of 
massive galaxies -- plays on star formation in low-mass galaxies 
\citep[see, e.g.][]{geha2012, carlsten2022}. 
\rrr{
We find that one UDG in our sample, AGC227965,
is quenched (SFR$<10^{-3}$ M$_\odot$ yr$^{-1}$) 
presumably due
to being a close satellite of MRK1324. We leave this galaxy in our
analysis as it still yields a significant \HI{} detection, but note that our discussion of the 
star formation models does not apply to this system due to the ongoing interaction. 
}
Secondly, and perhaps most importantly for 
the low surface brightness regime, an \HI-selected sample has redshift measurements from
the 21 cm line, which allows us to determine distances to relatively isolated UDGs.
This selection results in a total sample of 21 \HI-selected UDGs.

In order to make a fair assessment of the star formation activity of the UDGs, 
we also draw a reference sample of 32 NASA-Sloan Atlas (NSA) dwarfs at $d<120$ Mpc
with \HI{} measurements from \cite{bradford2015} to act as a 
``normal'' dwarf reference sample. For this work, we consider the 21
NSA dwarfs with stellar masses no greater than $\rm M_\star=3\times10^8\ M_\odot$ (the maximum stellar mass
covered by the UDG sample). 
This
results in a median stellar mass of 
$\langle \rm \log_{10}(M_\star/M_\odot)\rangle_{50}=8.0$ for the 
NSA sample and $\langle \rm \log_{10}(M_\star/M_\odot)\rangle_{50}=7.9$
for the UDG sample.

In \autoref{f:gascomparison} we show  
the galaxies in our sample
in stellar mass versus \HI{} mass (left), SFR versus \HI{} mass (center), and
the star-forming main sequence (SFMS, right). These results are obtained via the SED
fitting method presented in \citetalias{paperone} that we will summarize in 
\autoref{s:methods:sedfitting}.
In this figure and all subsequent figures, we show the NSA
sample in purple and the UDG sample in red. 
The results of the \cite{leroy2008} sample of nearby
galaxies and SDSS spectroscopic value-added catalog \citep{kauffmann2003, brinchmann2004, salim2007}
are shown in beige and grey, respectively. 
We find that the UDGs have high \HI{} masses for their 
stellar masses, and that they have low SFRs for their \HI{} masses.

\begin{figure*}[htb]
\centering     
\includegraphics[width=\linewidth]{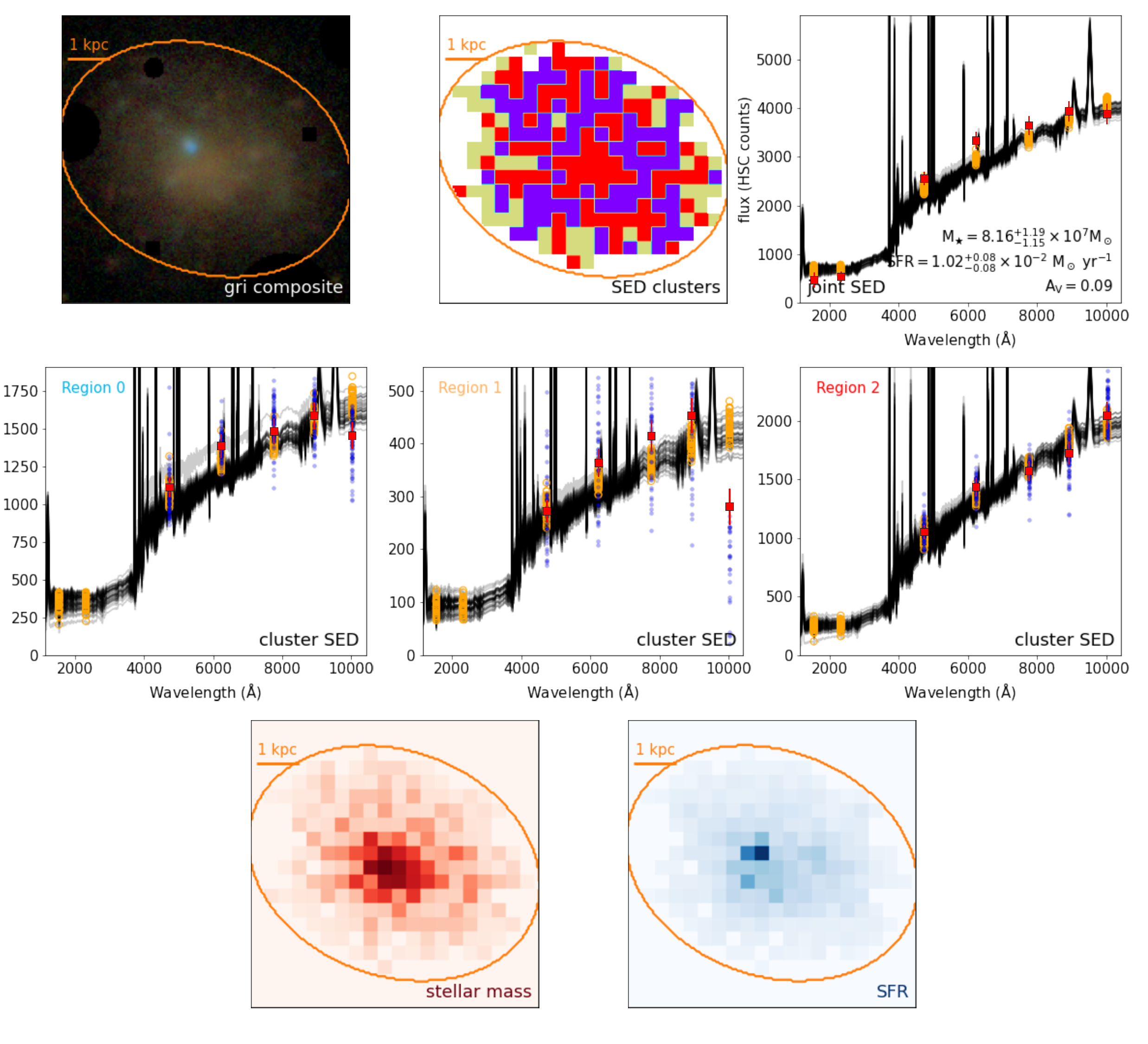}
\caption{ 
    An example of the fitting process for one of the UDGs (AGC 334349). 
    \textit{Top row:} from 
    left, the HSC $gri$-composite image, the SED cluster map, and the joint UV-optical SED fit. The joint model SED is
    the result of fitting the spatially resolved (cluster) optical photometry and global UV 
    photometry. The black curves show the model spectra, while the orange points show the
    model photometry for each filter. The red points show the global UV-optical photometry.
    The model fitting and figure fluxes are computed as HSC counts.
    \textit{Middle row:} optical-only results for the individual cluster SEDs. The 
    color of the title text corresponds to the color of the region in the top middle panel;
    the format is equivalent to the joint SED fit (top right), with the addition of the
    SEDs measured in individual spatial regions as blue points. 
    \textit{Bottom row:} the resultant stellar mass (left) and SFR (right) maps. 
    }
\label{f:janowiecki_example}
\end{figure*}

\subsection{SED Fitting}\label{s:methods:sedfitting}
In order to understand the star formation of UDGs down to 500 pc scales, we must
devise a method in which to measure their star formation on the relevant physical scales.
In the absence of spectroscopic or narrow-band measurements -- both of which are hindered,
but not rendered impossible, by the low surface brightness of the UDGs -- we turn instead 
towards a joint UV-optical SED fitting method that combines the star formation 
information from \galex{} FUV and NUV with the spatial resolution of the Hyper Suprime-Cam
Subaru Strategic Program (HSC-SSP) optical imaging, which attains a median seeing of 0.77''
in the g band \citep{aihara2019}. The \galex{} imaging is
over a factor of 6 lower in spatial resolution, with a
FUV and NUV PSF FWHM of 4.2'' and 4.9'', respectively. 
Convolving the HSC optical imaging to \galex{} seeing would
clearly constitute an unreasonable degradation of the
optical imaging. We have thus developed an SED fitting
process which jointly models global UV and spatially
resolved optical photometry.

A full description of the SED fitting process and validation can 
be found in \citetalias{paperone} and we direct the interested reader to this work.
We also present a schematic layout of the
procedure in \autoref{f:schematic}.

First, let us review the typical assumptions and parameters one must choose when
approaching an SED fitting problem.
We use a Kroupa initial mass function (IMF) throughout \citep{kroupa2001} with the
Flexible Stellar Population Synthesis (FSPS) library \citep{fsps_ref0, fsps_ref1}.
For this work, we hold stellar and gas-phase metallicity fixed at $Z=0.004$ (approximately
one-third solar). This choice is motivated by both technical and scientific concerns: first, due to the
joint nature of the SED fitting, we would like to minimize the number of parameters fit for 
each cluster SED (panel f, \autoref{f:schematic}). Second, the mass-metallicity relation is not well-understood
for these dwarf samples. Direct ($T_e$) measurements are limited to nearby ($d\lesssim20$ Mpc) samples of
typically ``normal'' dwarfs \citep{lee2006, berg2012, jimmy2015}; these direct measurements can show 
significant offsets relative to each other and to SED fitting methods used for more distant samples
\citep{bellstedt2021}. We thus choose to adopt a typical value following the 
literature compilation of \citep{bellstedt2021} for a galaxy of $M_\star\sim 10^8 M_\odot$.
\rrr{We find in \citetalias{paperone} that our results are not significantly impacted by our assumption
of a fixed metallicity; assuming solar metallicity shifts the estimated stellar mass and SFR of our 
dwarf samples by $\sim 0.08$ dex on average. The star formation rate surface densities and stellar mass
surface densities estimated using solar metallicity models are shifted on by less than 0.01 dex
on average compared to their fiducial values.
}

We also adopt an exponentially declining star formation history 
\begin{equation}\label{e:expdeclining}
    {\rm SFR} (t) \propto e^{-(t-t_0)/\tau}
\end{equation}
for all of the models in this work.

Our SED fitting method uses the \textsf{emcee} implementation of the
Markov Chain Monte Carlo method for parameter estimation \citep{foremanmackey2013}. We
begin with a joint fit to the UV and optical global photometry; this allows us to  
estimate the reddening in the galaxy, and provides a reasonable point at which to 
instantiate walkers in the joint fitting step. We then subdivide the galaxy into 
regions no smaller than twice the full width at half-maximum (FWHM) of the PSF in our
PSF-matched optical images. These regions are required to have a median signal-to-noise
ratio of 3 across the five optical bands. Then, we use K-means clustering in the space of optical photometry to 
assign these regions to clusters that have similar SEDs (\autoref{f:janowiecki_example}, top middle panel)
-- this reduces the number of
individual SEDs that we must fit in the joint modeling step. Finally, we fit the 
global UV data jointly with the cluster optical SEDs
(\autoref{f:janowiecki_example}, middle row): that is, we 
adopt the likelihood:
\begin{equation}\label{e:jointlikelihood}
\begin{split}
\log \mathcal{L} \propto& -\frac{1}{2} \sum_{i \in UV} \frac{( \sum_j^N f_{i,j} - \sum_j^N \hat f_{i,j} )^2 }{\sum_j^N \sigma_{f_{i,j}}^2} + \log(\sum_j^N \sigma_{f_{i,j}}^2)\\
& + -\frac{1}{2} \sum_j^N \sum_{i\in Opt}  \left[ \frac{( f_{i,j} - \hat f_{i,j} )^2 }{\sigma_{f_{i,j}}^2} + \log(\sigma_{f_{i,j}}^2) \right],
\end{split}
\end{equation}
where $i$ refers to the bandpass index and $j$ refers to the cluster index. This allows us
to jointly fit the spatially resolved optical photometry and the global, unresolved UV 
photometry. From this inference, we can then immediately extricate the stellar mass
surface density, given the amplitude of the cluster SEDs,
as well as the star formation rate surface density, which we estimate by integrating the
inferred star formation history over the past 100 Myr (\autoref{f:janowiecki_example},
bottom row). We choose 
100 Myr given that due to our UV coverage, 
we should be most sensitive to star formation averaged over 
this timescale. 

We finally compute the average stellar mass density and star formation
rate surface density over 500 pc, 1 kpc, and global scales. In this step, we 
require that the averaging scale is larger than the 
scale of the regions in
which we measure the region SEDs (panel e of \autoref{f:schematic}). 
In practice, 
this affects only five galaxies: 
four are UDGs (AGC 227965, AGC 322019, AGC 198543,
and AGC 238961), and one is a NSA dwarf (NSA ID 17750). The results of these galaxies
are thus shown on 1 kpc and global scales, but not on 500 pc scales -- their 
exclusion at 500 pc does not constitute a statistically significant impact on our
analysis.

\section{Background on PRFM theory}\label{s:prfm}
We will 
summarize the basic points of  
pressure-regulated, feedback-modulated (PRFM) star formation, but 
direct the reader to \citeok{} for an in-depth discussion of the theory.

For a galaxy disk in a quasi-steady state or in 
vertical dynamical equilibrium, the weight (per unit area) of 
the ISM should be balanced by the pressure difference
between the midplane and the top of the gas layer. Since
the pressure is generally decreasing rapidly along the
z-axis (perpendicular to the disk plane), the 
midplane pressure must match the overlying weight.
The weight
of the ISM, $\mathcal{W}$, 
is a sum of the contribution by the gas and by the external
components (stars, dark matter), 
\begin{equation}
\begin{split}
    \mathcal{W} &= \int_0^{z_{\rm max}} dz\ \rho(g_{\rm gas} + g_{\rm ext})\\
                &= \mathcal{W}_{\rm gas} +  \mathcal{W}_{\rm ext},
\end{split}    
\end{equation}
where $z_{\rm max}$ refers to the vertical confine of the gas disk, 
$g_{\rm gas}$ and $g_{\rm ext}$ are the vertical components of the
gravitational field due to the gas and the external gravitational 
potential, and $\mathcal{W}_{\rm gas}$ and $\mathcal{W}_{\rm ext}$ are
the corresponding weight components.

Assuming slab geometry, 
it can be shown that the weight from the gas gravity is
$\mathcal{W}_{\rm gas} = (\pi G \Sigma_{\rm gas}^2)/2$, where 
$\Sigma_{\rm gas}$ is the total gas surface density.
If the external gravity dominates, we can approximate the weight 
as $\mathcal{W}_{\rm ext} = \Sigma_{\rm gas} \sqrt{2 G \rho_{\rm sd}} \sigma_{\rm eff}$, where $\rho_{\rm sd}$ is the combined density of
stars and dark matter and $\sigma_{\rm eff}$ is the effective vertical velocity dispersion of the gas. This effective velocity dispersion includes the turbulent, thermal, and magnetic contributions. To estimate the weight from observables,
we take a simplified form called the dynamical equilibrium pressure \pde{} that is obtained by combining the two
weight contributions as introduced in \citealtok{}:
\begin{equation}\label{e:weightpde}
    \begin{split}
        \mathcal{W}&\approx P_{\rm DE} = \frac{\pi G \Sigma_{\rm HI}^2}{2}  +  \Sigma_{\rm HI}\sqrt{2 G \rho_{sd}}\sigma_{\rm eff}. \\
    \end{split}    
\end{equation}
This approximation is good to within $\approx 20$\%
(see \citealtok{} for a more complete derivation of the
full weight expression), and has been adopted in the
literature to estimate the total weight from observable
quantities \citep[see, e.g.][]{sun2020}.
The reader will note that
we have made a notable change to the construction of $\mathcal{W}$. 
We do not have molecular gas estimates for our samples, and thus 
take the base assumption that these galaxies are \HI-dominated and 
that the molecular gas is a minor contributor to the total gas mass of
the ISM. That is, we assume $\Sigma_{\rm gas}\approx \Sigma_{\rm HI}$.
We assume a $\sigma_{\rm eff}=10 \kms{}$, as in 
previous works \citep[see, e.g.][]{sun2020}. 
This assumption is reasonable for the dwarf samples
at hand as $\sigma_{\rm eff}$ is likely dominated by the
warm gas sound speed. 


Having laid out an observational estimate of the weight (that is, the dynamical equilibrium pressure), which is an estimate of the
total midplane pressure, we now want to make a connection 
between the total pressure and star formation activity.
Total pressure here is defined as the
sum of contributions from thermal pressure, turbulent pressure (vertical Reynolds stress), and
vertical Maxwell stress as
$P_{\rm tot} = P_{\rm th} + P_{\rm turb} + \Pi_{\rm mag}$. Each of these pressure terms is expected to be linked to 
star formation activity: thermal pressure scales largely with the mean 
FUV field intensity, turbulent
pressure is sourced by momentum injection from supernovae, and magnetic pressure -- though relatively
less well understood -- is thought to scale to some degree with
turbulent kinetic pressure via galactic dynamo.

A great deal of work has been devoted to 
quantifying the relationship between these pressure terms and star formation
activity. The key parameter quantifying this relation is the
``feedback yield'', defined as
\begin{equation}
    \Upsilon_{\rm X} \equiv \frac{P_{\rm X}}{\Sigma_{\rm SFR}}
\end{equation}
where X refers to a given pressure component.
In particular, \cite{ostriker2011} showed that the turbulent pressure is related to
SFR surface density as $P_{\rm turb}= p_\star/(4 m_\star) \Sigma_{\rm SFR}$, where
$p_\star$ is the spherical momentum injection per supernova and $m_\star$ is the
total mass of stars formed per star that will become a supernova. For a typical IMF \citep[e.g.,][]{kroupa2001}, $m_\star\approx100M_\odot$.
Recently, 3D supernova-driven bubble expansion simulations have
converged to a characteristic value of 
$p_\star/m_\star \sim 1000-3000 \kms{}$, giving rise to $\Upsilon_{\rm turb} \sim 250 - 750 \kms{}$. This value is
insensitive to the background density and metallicity
\citep[e.g.][]{kim2015, kim2017, martizzi2015, iffrig2015, fielding2018, oku2022}

Turning our attention to the thermal pressure,
it is established that $P_{\rm th}$ should
scale with the FUV field intensity as the photoelectric heating \citep[e.g.,][]{bakes1994,weingartner2001} is the dominant heating source, which is dependent on both \sigsfr{} and
attenuation from the ISM. In particular, a scaling relation can be written
for $\Upsilon_{\rm th}$ with respect to solar neighborhood conditions as:
\begin{equation}
\begin{split}
    \Upsilon_{\rm th}(f_\tau, \Sigma_{\rm HI}, Z_d') =& 240 \kms{}\ \times \\
    & \frac{4.1 f_{\tau}/f_{\tau,\odot}}{1 + 3.1\left(\frac{\Sigma_{\rm HI}Z_d' f_{\tau}/f_{\tau,\odot}}{10\ {\rm M_\odot}\ {\rm pc}^{-2}}\right)^{0.4}},
\end{split}    
\end{equation}
where we assume a dust metallicity of $Z_d'=Z/Z_\odot=1/3$, in line with our
SED fitting assumptions. We also introduce
$f_\tau$, which is the mean attenuation factor of the UV radiation field. Under the assumption
of slab geometry and uniform gas and source distribution, the radiation transfer solution is
\begin{equation}\label{e:ftau}
    f_\tau \equiv \frac{1 - E_2(\tau_{\rm FUV}/2)}{\tau_{\rm FUV}},
\end{equation}
where $E_2(n)$ is the second exponential integral and $\tau_{\rm FUV}$ is the mean FUV optical depth in the direction perpendicular to the disk. We estimate  $\tau_{\rm FUV}$ for our galaxies by computing $A_{\rm FUV}$ assuming a 
\cite{calzetti2013} extinction curve and the $A_V$ inferred from our 
global SED fits. 
\cite{ostriker2010} showed that 
for solar neighborhood-like conditions one arrives at 
$\Upsilon_{\rm th}\approx240 \kms{}$
\citep[see Equation 15 of ][or, equivalently, Equation 12 of \citealtok{}]{ostriker2010}. 
We compute somewhat higher values of $\Upsilon_{\rm th}$ for the present samples due to low $\Sigma$ and low $Z'$ than those of solar neighborhood, with a
median value of $\Upsilon_{\rm th}\sim 660\kms{}$ at 1 kpc scales.

It is generally expected that the saturation level of the magnetic stress is set by a fraction of the turbulent stress (e.g., $\Upsilon_{\rm mag} \sim 0.5 - 1.0 \Upsilon_{\rm turb}$; \citealt{kim2015}). However, the quantitative prediction may still depend on the details of galactic dynamo.
In this
work, we adopt a fiducial $\Upsilon_{\rm turb}=500 \kms{}$ and a scaling of 
$\Upsilon_{\rm mag} = 0.75\Upsilon_{\rm turb}$. 

We thus arrive at a 
theoretical PRFM prediction of
\begin{equation}\label{e:Yfixed}
    \Upsilon_{\rm tot} = \frac{P_{\rm tot}}{\Sigma_{\rm SFR}} \sim 875\kms{} + \Upsilon_{\rm th}(f_\tau, \Sigma_{\rm HI}).
\end{equation}
Numerical results from the TIGRESS\footnote{Three-phase Interstellar medium in Galaxies Resolving 
Evolution with Star formation and Supernova feedback} framework \citep{tigressintro}
validate the theoretical assumptions of the PRFM theory and calibrate $\Upsilon_{\rm tot}$ as a function of $P_{\rm DE}$ as (\citealtok{}, 
Equation 25c):
\begin{equation}\label{e:Y25c}
    \log_{10}\left(\frac{\Upsilon_{\rm tot}}{\kms{}}\right) = -0.212\log_{10}\left(\frac{P_{\rm DE}}{k_B\ {\rm cm}^{-3}\ K}\right) + 3.86.
\end{equation}

Given a method to compute \Ytot{} (we will use the numerical prescription of 
\autoref{e:Y25c}, but will include a comparison to \autoref{e:Yfixed} 
in \autoref{s:results:empiricalYtot}), the PRFM prediction for SFR surface density is
\begin{equation}
    \Sigma_{\rm SFR} = \frac{P_{\rm DE}}{\Upsilon_{\rm tot}},
\end{equation}
where the right hand is composed of our observable quantity (\pde{}) and  
the theoretical/numerical prediction (\Ytot{}).

From here,
we will consider whether the markedly low SFE(\HI) of our observed UDG sample
can be explained within the framework of PRFM star formation.

\section{Results}\label{s:results}\label{s:results:sftheory}
Having provided the reader a brief introduction to PRFM theory, we may now proceed
to the application of this framework to the present sample.
Before we jump headlong into the computation,
however, it is of substantial importance to first justify that 
the PRFM theory of star formation is applicable to our dwarf samples, and how we will 
go about estimating the parameters necessary to predict star formation within the 
model framework.

\subsection{The validity of PRFM assumptions in the present sample}
A preliminary question that we must first address before applying 
the PRFM model to the
sample at hand is whether the equilibrium disk assumptions that lie at the
heart of the theory
are satisfied in our low-mass systems. In this work, we are averaging
both spatially (over at least 500 pc) and temporally (over approximately
100 Myr, given that \galex{} FUV is included in the SED fitting). 
The equilibrium assumption is valid if either the averaging timescale
or length scale is large enough to average over deviations from 
equilibrium. 
In this section, we will provide an argument that 
our averaging timescale is long enough to justify the equilibrium
assumption, but we note that the spatial averaging is also 
likely sufficient to validate the equilibrium assumption.
Indeed, it should be noted that
spatial averaging allows PRFM theory to be used for 
star formation tracers that track SFR on shorter timescales 
than our time-averaging
arguments may otherwise imply (e.g. H$\alpha$, which
traces star formation activity on $\sim\! 10$ Myr timescales). 

Before moving onto these timescale arguments, however, it is worth
considering whether these dwarfs are well-described by 
gaseous disks. There is 
substantial evidence from \HI{} rotation curves that dwarfs (including UDGs) in this mass range do host
\HI{} disks \citep{hunter2012, greco2018b, mancerapina2019, mancerapina2020, gault2021, mancerapina2022}.
There are, to the authors' knowledge, no published analyses of the stellar kinematics of field 
UDGs, but there
is evidence that UDGs in the field are ``puffy'' but largely axisymmetric systems characterized
by a mean disk thickness of $ \sim 0.5 R_{\rm d}$ where $R_{\rm d}$ is the 
disk length \citep{kadofong2021a}. 
We thus generally find long-lived disks for the galaxies in our sample so that the slab geometry assumption of \oten{} is reasonable for the present work. We then must carefully consider the assumptions of thermal and dynamical equilibrium due to proposals that dwarfs have generally bursty star formation histories, and that UDGs in particular may be formed
via particularly bursty histories \citep{chan2018, dicintio2017}. 

The thermal equilibrium assumption of the \oten{} model may be disrupted if the cooling and heating time scales are so long that the majority gas is in out of equilibrium state. In this case, the gas would not promptly respond to the change in the heating rate and hence the star formation rate.
We can estimate the validity of this assumption by comparing the time-scale over which we measure the star formation rate, to the cooling and heating time for these dwarfs, which quantifies the time it takes to re-establish a thermal equilibrium.
In particular, let us write the
cooling and heating time of warm gas as:
{}
\begin{equation}
    t_{\rm cool} = \frac{k_B T_w}{n \Lambda} = \frac{k_B T_w}{\Gamma},
\end{equation}
where $k_B$ is Boltzmann's constant, $T_w \sim 5000-8000$ K is the temperature of the warm gas, $n\Lambda$ and $\Gamma$ 
 are the cooling and heating rate per particle, respectively.
With the $\Gamma \sim 5 \times 10^{-27} - 10^{-26}\, {\rm erg/s}$, appropriate for low-metallicity gas \citep{wolfire2003},
we have a cooling time of $t_{\rm cool}\sim 2-6$ Myr.
This cooling time is significantly shorter than the
timescale over which we measure our star-formation rates (100 Myr) -- thus, 
the time scale over which our measurement is averaging is larger 
than the cooling time by roughly an order of magnitude. This indicates that 
even the smallest regions over which we are averaging (500 pc) should follow
the equilibrium expectation.

Variations in star formation can also drive large-scale changes in the dynamical equilibrium structure of the galaxy \citep{elbadry2016, orr2019}. We can quantify the degree to which dynamical equilibrium is a valid assumption for the system under consideration here by comparing the time-scale over which we measure the SFR to the vertical crossing time of the disks, which generally quantifies the time over which excess kinetic energy is dissipated in the galaxy \citep{ostriker2001}. Generally, we can write $t_{\rm cross} \approx H_{\rm gas}/\sigma_{\rm eff}$, where $H_{\rm gas}$ is the gas scale height. 
We estimate the gas scale height as
\begin{equation}
\begin{split}
    H_{\rm gas} = \frac{2\sigma_{\rm eff}^2}{\pi G \Sigma_{\rm HI} + 2 \sigma_{\rm eff}\sqrt{2G\rho_{sd}}} ,
\end{split}
\end{equation}
which is equivalent to equation 5 of \cite{ostriker2022}
except that we assume $\Sigma_{\rm gas}\approx \Sigma_{\rm HI}$ and
take $W\approx P_{\rm DE}$. 
This estimate results in 
 $t_{\rm cross} \approx 30\, {\rm Myr}$ for the samples
 at hand -- again, significantly smaller than the timescale over
 which we are measuring star formation activity.


\subsection{Estimating $\rho_{\rm sd}$}\label{s:results:estimatingrhosd}
Having addressed these assumptions, we now proceed 
to the sample at hand. In order to estimate \pde{} (\autoref{e:weightpde}), we will
need an estimate for the midplane density of stars ($\rho_\star$) 
and dark matter ($\rho_{\rm dm}$), 
$\rho_{\rm sd} = \rho_\star + \rho_{\rm dm}$.
We estimate the stellar mass density at the midplane as
$\rho_\star = \Sigma_\star / (2H_\star)$, where $H_\star$ is the disk scale height and $\Sigma_\star$
is the stellar mass under the assumption of an exponential density profile. 

\begin{figure*}[t]
\centering     
\includegraphics[width=\linewidth]{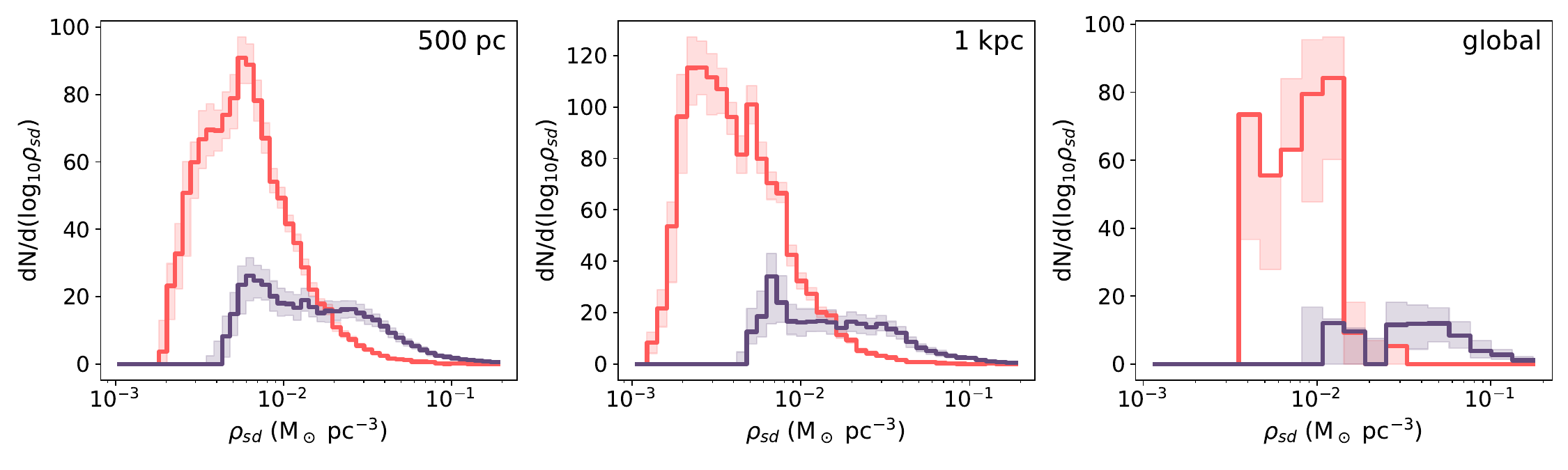}
\caption{ 
    The distribution over $\rho_{sd}$ for our fiducial 
    dark matter halo assumptions (see text) for the NSA dwarfs (purple) and 
    UDGs (red) at 500 pc, 1 kpc, and global scales from left to right. 
    The shaded regions are bounded by the 16\thh{} to 84\thh{} percentiles
    of the distribution. The NSA dwarfs are 
    characterized by systematically higher stellar$+$dark
    matter densities, which is unsurprising given that the UDGs are 
    characterized by low stellar surface densities.
    }
\label{f:rhosddist}
\end{figure*} 

We obtain this estimate statistically
by using the measured scale lengths of our galaxies in conjunction with the
three-dimensional shape distributions inferred by \cite{carlsten2021} 
for the NSA dwarf
sample and by \cite{kadofong2021a} for the UDG sample. 
We assume in both cases that the three-dimensional
shape distribution of the galaxy sample measured at one effective 
radius is the same as the distribution measured
at the scale length $l_\star$. 
That is to say that we assume $\camath{} \approx H_\star/l_\star$
where $\camath{}$ indicates the ratio between the smallest and largest principle axes of the ellipsoid that describes
the galaxy shape distribution at one effective radius, and
$H_\star$ and $l_\star$ are the stellar scale height and length, 
respectively. 
Both the UDGs \citep{kadofong2021a}
and the ``normal'' dwarfs \citep{kadofong2020c} are well-characterized
by exponential surface brightness
profiles, meaning that \reff{}$\approx 0.6 l_\star$ assuming a constant
stellar mass-to-light ratio. 
We can then estimate the probability distribution function of $H_\star$ as:
\begin{equation}
    \begin{split}
    {\rm Pr}[H_\star] =& (2\pi\sigma_{\camath{}}^2)^{-1/2} \times \\
    &\exp \left[ \frac{-(H_\star/l_\star - \mu_{\camath{}})^2}{ (2\sigma_{\camath{}}^2)} \right]
    \end{split}
\end{equation}
where
$\mu_{\camath{}}$ and $\sigma_{\camath{}}$ are the inferred mean and standard deviation of the bivariate 
Gaussian used to describe that intrinsic shape distribution in \cite{carlsten2021} and \cite{kadofong2021a}.

We use the \cite{carlsten2021} 3D shapes
because of the overlap in stellar mass between the two samples, but we 
note that their sample is comprised only of satellite galaxies. 
We therefore also compute
\sigsfr{} predictions using the 3D intrinsic shape distribution of \cite{kadofong2020c},
which is incomplete at the relevant stellar masses but includes field galaxies,
and find no 
significant difference in the SFR predictions between the assumptions of intrinsic shape distribution.

We opt
to not implement inclination corrections for our sample because the stellar disks are not
thin and because empirical results suggest that there is significant variation in the 
three-dimensional shapes and thicknesses of the dwarf stellar disks 
\citep{kadofong2020c, carlsten2021}; however, we compute inclination estimates using the mean
disk height-to-length ratio following \cite{holmberg1958} and find that our results would not
qualitatively change if we did implement such an inclination correction. \rrr{We note that 
adopting an inclination correction would systematically lower the stellar mass surface density
and star formation rate surface density estimates in this work by an average of $\sim 34\%$
below fiducial values, though we stress that this value is highly uncertain due to the assumptions
made in the application of the \cite{holmberg1958} correction. Furthermore, because the 
feedback yield \Ytot{} is the ratio between the midplane pressure and the star formation rate
surface density, the effect of an inclination correction applied to both the stellar mass
and star formation rate surface density estimates should be lessened.}

\begin{figure*}[t]
\centering     
\includegraphics[width=\linewidth]{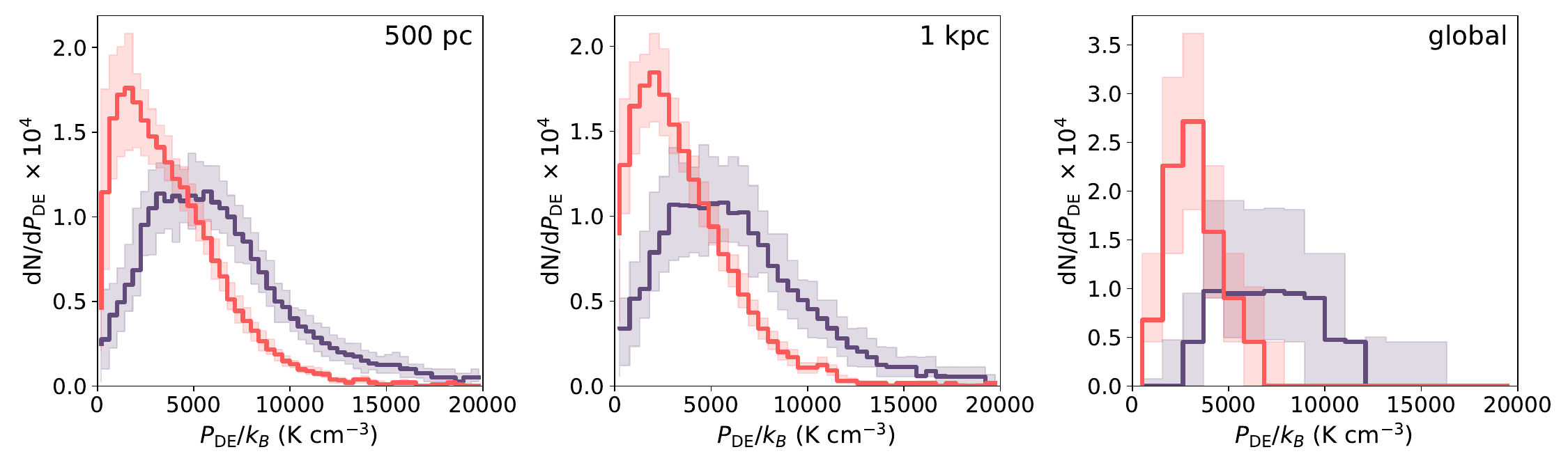}
\caption{ 
    The distribution over \pde{} for the NSA dwarfs (purple) and 
    UDGs (red) at 500 pc, 1 kpc, and global scales from left to right. 
    The shaded regions 
    at 500 pc and 1 kpc are bounded by the 16\thh{} to 84\thh{} percentiles
    of the distribution. We find no significant variation in \pde{} as a function
    of spatial scale, but find that
    the NSA dwarfs have systematically higher
    dynamical equilibrium pressures than the UDG dwarfs.
    }
\label{f:pdedist}
\end{figure*}

We now move on to the estimation of the dark matter density, $\rho_{dm}$. Here, we must adopt a 
dark matter halo profile for each galaxy. This component is by far the most
uncertain ingredient in the star formation prescription -- not only do we not have
any constraint on the individual dark matter halos in which these galaxies live, the
overall stellar-to-halo mass relation is highly uncertain for low-mass galaxies.
In order to take into account the
scatter and uncertainty in the stellar-to-halo mass relation at these masses,
we compute the predicted star formation rate using three different dark matter
halos: a fiducial best-guess halo mass using the stellar-to-halo mass relation of the FIRE-2 simulations \citep[which predicts halo masses 
between $1.6\times10^{10}$ and $6\times10^{10}$ M$_\odot$ for the 
sample at hand, ][]{hopkins2018}, 
a halo with $ M_h = 10^{12} M_\odot$, and  a halo with $ M_h = 10^{9} M_\odot$.
We assume an NFW profile for all dark matter halos in this work \citep{NFW}.
We assign a concentration based on the concentration-mass
relation of \cite{child2018}. We assign the uncertainty due to the dark matter
density to be the difference between $\rho_{dm}$ estimated for
a very massive ($ M_h = 10^{12} M_\odot$) and very 
low-mass ($ M_h = 10^{9} M_\odot$)
halo, as it is unlikely that any of our galaxies are living in halos with a stellar-to-halo mass ratio more extreme
than 0.1 (in the case of the extreme low-mass halo) or 0.0001 (in the case of the extreme
high-mass halo). For our fiducial halo assumptions, the 
combined stellar and dark matter density tends to be 
dominated by dark matter, as is expected for
these low-mass galaxies \citep[see, e.g.][]{oh2011}.
The UDGs tend to be more dark matter-dominated, with a 
median $\rho_\star / \rho_{sd} = 0.14$ compared to the
mildly dark matter-dominated NSA dwarfs (median
$\rho_\star/\rho_{sd} = 0.46$). 

We show the distribution over $\rho_{sd}$ for our fiducial dark matter halo 
assumption in \autoref{f:rhosddist}. We find that the NSA dwarfs are characterized
by systematically higher values of $\rho_{sd}$, which is unsurprising given
that the UDGs are characterized by relatively low stellar surface densities. 

\subsection{\pde{} Estimates}
Having estimated both $\Sigma_{\rm HI}$ and $\rho_{sd}$, we may now arrive at an
estimate of the dynamical equilibrium pressure (\pde{}). We compute this quantity
following \autoref{e:weightpde}.

Before considering the full PRFM prediction, it is of interest to first consider
how \pde{} varies between the NSA and UDG samples. We show the overall distribution
of \pde{} at each spatial scale used in this work (500 pc, 1 kpc, and global) 
for each sample in \autoref{f:pdedist}. We find that there is no significant variation
in \pde{} distribution as a function of spatial scale, but that the UDGs tend to
have lower dynamical equilibrium pressures than do the NSA dwarfs. This result 
is as expected due to the low $\rho_{sd}$ values of the
UDGs (see \autoref{f:rhosddist}). The relatively low stellar$+$dark matter
densities (and relatively similar \HI{} surface densities) of the UDGs 
also imply that $\mathcal{W}_{\rm gas}$ is a larger contributor to $\mathcal{W}$
in the UDGs than it is in the NSA dwarfs. 

\begin{figure*}[t!]
\centering     
\includegraphics[width=\linewidth]{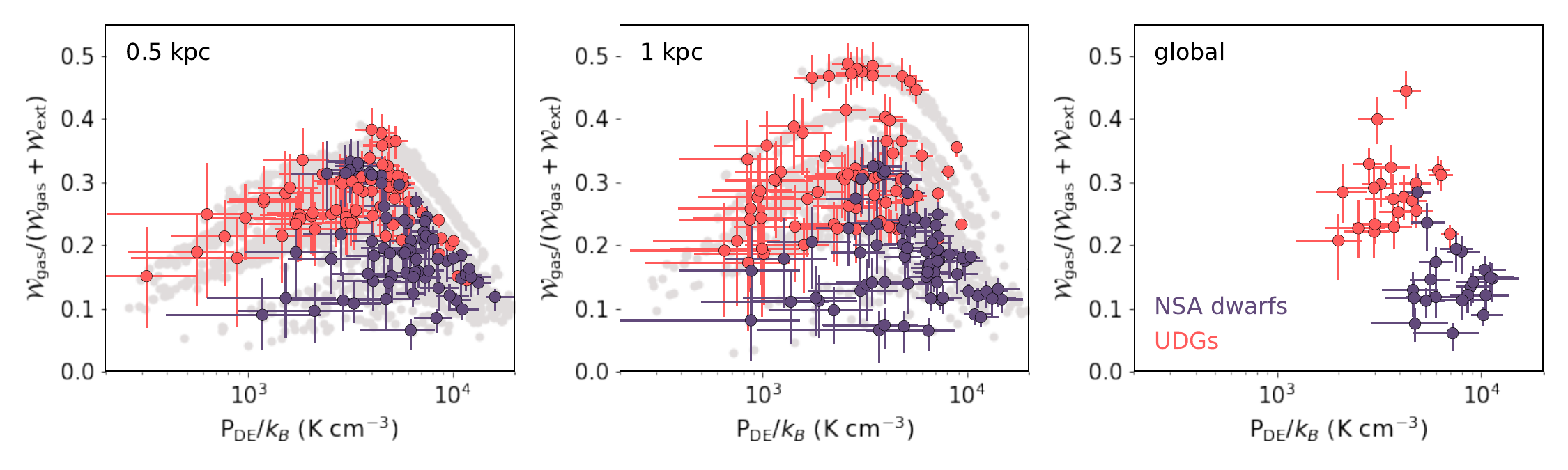}
\caption{ 
    The contribution of self-gravity to the total midplane weight
    ($\rm \mathcal{W}_{gas} / \mathcal{W}$) as a function of dynamical equilibrium
    pressure ($P_{\rm DE}$) as a function of spatial scale (500 pc to global average,
    from left to right). We show individual errorbars for a random sample of
    regions from the UDG sample (red) and the NSA sample (purple); the full sample 
    is shown in grey. Note that the tracks visible in the left and middle panels are
    a result of the median assumed \HI{} profiles; the uncertainty in the
    \HI{} profiles result in a larger dispersion in $\rm \mathcal{W}_{gas} / \mathcal{W}$ than is implied by the tracks alone (as is shown by the errorbars
    of the colored points). 
    We find that the external weight ($\mathcal{W}_{\rm ext}$) is 
    relatively less important in the UDGs as compared to the NSA dwarfs.
    }
\label{f:wratio}
\end{figure*}

We show 
$\mathcal{W_{\rm gas}}/\mathcal{W}$ in \autoref{f:wratio} as computed from
\autoref{e:weightpde} where $\mathcal{W}\approx P_{\rm DE}$ and
$\mathcal{W}_{\rm gas}= (\pi G \Sigma_{\rm HI}^2)/2$. 
Each panel shows the dynamical equilibrium pressure  versus the fraction of
$\mathcal{W}$ contributed by $\mathcal{W}_{\rm gas}$. From left to right, the 
panels show measurements at 500 pc, 1 kpc, and global scales.
For visual clarify, we only show errorbars for a random subset of the 
regions measured in this work. The colored points show 
a random sample of UDG regions (red) and NSA regions (purple) with their errorbars
included; the full sample is shown by the grey scatter. The tracks visible in 
the left and middle panels are a result of the median \HI{} profiles we assume for the
galaxies; the uncertainty in these profiles result in a wider dispersion than is 
implied by the tracks (as shown by the vertical extent of the errorbars).
Indeed, we find that
$\mathcal{W}_{\rm gas}$ accounts for a larger fraction of the total weight in the 
UDGs than it does in the NSA dwarfs.

\subsection{PRFM predictions}
With the assumptions validated and parameters estimated, we may now move on to predicting \sigsfr{}
from the PRFM framework for our observed galaxies. 
We base our predictions upon the 
numerically estimated form of \Ytot{} presented by 
\citeok{} using the TIGRESS simulations, as discussed in
\autoref{e:Y25c}.

\begin{figure*}[htb]
\centering     
\includegraphics[width=\linewidth]{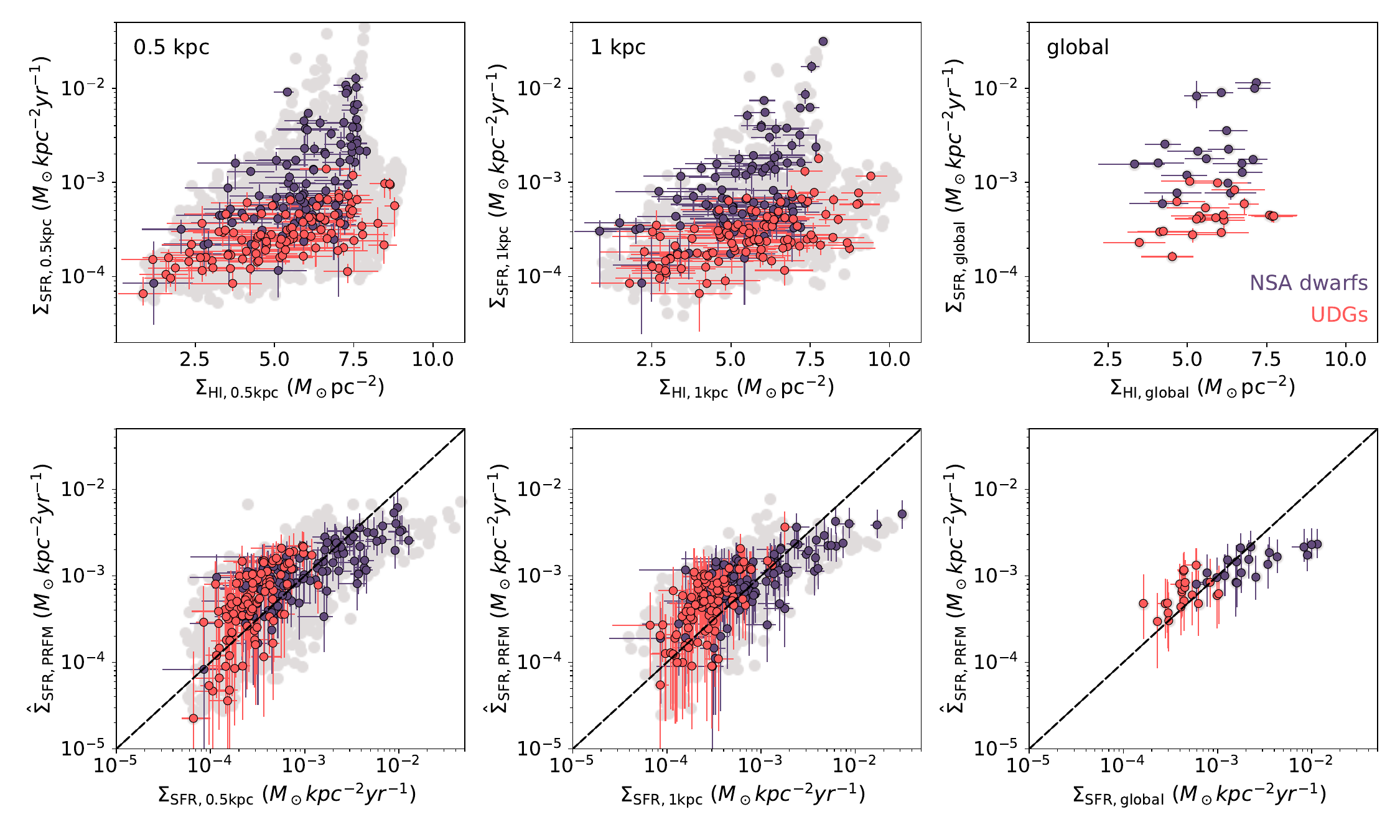}
\caption{ 
    \textit{Top Row:} 
     a comparison between \HI{} surface density estimates and
    SFR surface density measurements at 500 pc (left column), 1 kpc (middle
    column), and global scales (right column).  In all panels, we only show NSA galaxies that are within the stellar
    mass range of the UDG sample. For visual clarity, we show a random subset of (at most) 
    70 points for each sample with individual errorbars and colored by 
    sample source; we show the full sample as the underlying grey scatter.
    \textit{Bottom row:} a comparison between the observed SFR surface densities 
    and the PRFM predictions. Again,
    red points show UDGs while purple points show NSA dwarfs.
    }
\label{f:PRFMcomparison}
\end{figure*}

In the top row of \autoref{f:PRFMcomparison}, we show the relationship between \sigsfr{} and \HI{} surface
density for our NSA dwarfs (purple) and UDGs (orange) averaging over 500 pc regions, 1 kpc regions, and the
full area of the galaxy. 
As was shown in more detail in \citetalias{paperone}, the UDGs
form stars at consistently lower efficiencies (relative to their \HI{} content) relative to the NSA dwarfs. 

We then show the relation between the observed and 
predicted star formation rate surface densities in
the bottom row of \autoref{f:PRFMcomparison} 
for 500 pc, 1 kpc, and global averages. 
We find that 
the \oten{} prescription succeeds in reproducing the 
star formation rate surface density measurements of the UDGs
at all spatial scales probed, which we will explore quantitatively
via a comparison between the predicted and measured values of \Ytot{} in
the following section.
The prescription under-predicts $\Sigma_{\rm SFR}$ for the 
most vigorously star-forming regions of the NSA sample -- this under-prediction can be understood if
one considers that the NSA dwarfs may have considerable stores of molecular hydrogen 
\citep[see, e.g.][]{leroy2008, delosreyes2019}. The 
discrepancy is most obvious in the globally averaged estimates, which is due to
the global averages acting as a $\Sigma_{\rm SFR}$-weighted average (where the 
most vigorously star-forming regions dominate the total mean signal). We also see that the \HI-only PRFM predictions perform
best for the NSA dwarfs at 500 pc scales, which is consistent 
with the picture where
much of the galaxy is dominated by atomic hydrogen (over molecular hydrogen as
has been suggested previously in, e.g. \citealt{leroy2008}).
Nonetheless, we find that PRFM theory well-describes the majority of the 
regions within both galaxy samples and thus naturally predicts the low SFE(\HI)
of the \HI-rich UDGs as a consequence of their low 
ISM weight.

\subsection{Empirical estimates of \Ytot{}}\label{s:results:empiricalYtot}
Finally, though we have adopted \Ytot{} as a way to predict 
\sigsfr{}, it is also informative 
to re-frame the computation as a comparison between the theoretically predicted
values for \Ytot{} (\autoref{e:Yfixed} and \autoref{e:Y25c}) and 
empirically-derived \Ytot{} as the ratio between our observed \sigsfr{} 
measurements and \pde{} estimates. In this way, we can quantitatively
compare the 
relationship between star formation and dynamical equilibrium
pressure in our dwarf samples and similar measurements of more massive
galaxy samples from the literature within the context of theoretical
expectations.

Because the theoretical 
\Ytot{} depends explicitly upon $\tau_{\rm FUV}$ 
(\autoref{e:ftau}), we compute a running median predicted 
\Ytot{} for our sample as a function of \pde{} in bins of 
width 0.1 dex. The shaded region in each panel spans the
16\thh{} to 84\thh{} percentiles of the theoretical \Ytot{}
prediction over the same domain.

\begin{figure*}[htb]
\centering     
\includegraphics[width=\linewidth]{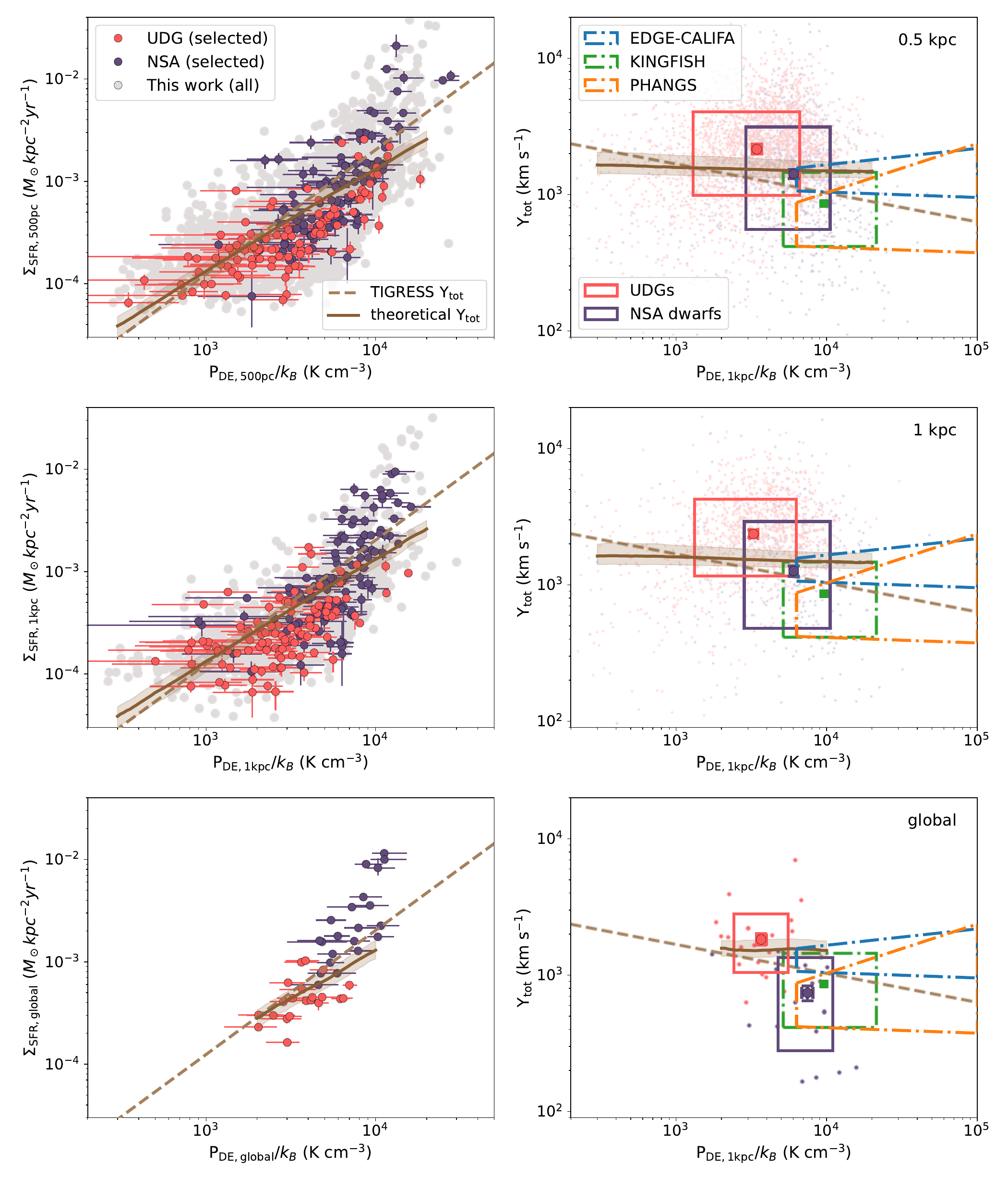}
\caption{ 
    \textit{Left:} an estimate of the dynamical equilibrium pressure
    ($P_{DE}$) for our sample.
    As in \autoref{f:PRFMcomparison}, we show a random sample of
    points with their individual errorbars and colors (red for the 
    UDG sample, purple for the NSA sample) and the full sample in
    grey. We overplot the theoretical (\autoref{e:Yfixed}) and
    numerical (\autoref{e:Y25c}) predictions of
    \citeok{} as the dashed and solid lines, respectively.
    \textit{Right:} an empirical estimate of \Ytot{} 
    ($=P_{\rm DE,obs}/\Sigma_{\rm SFR,obs}$) as a function of dynamical equilibrium
    pressure.
    We compare to kpc-scale observations from PHANGS 
    \citep[orange dot-dashed,][]{sun2020},
    EDGE-CALIFA \citep[blue dot-dashed,][]{barreraballesteros2021}, 
    and KINGFISH \citep[green dot-dashed,][]{herreracamus2017}, as well
    as to the theoretical predictions of \citeok{}.
    }
\label{f:PDEYtot}
\end{figure*}

We show such a comparison in \autoref{f:PDEYtot}. At left, we show \sigsfr{} 
versus \pde{} at 500 pc, 1 kpc, and global scales with the predictions using \Ytot{} 
over-plotted
(\autoref{e:Yfixed} solid, \autoref{e:Y25c} dashed). At right we show the
empirical estimate for \Ytot{} as a function of \pde{}, again with both 
the theoretical and numerical predictions for \Ytot{} shown in brown. Here
we show the median value of \Ytot{} for the UDG and NSA samples as the 
red and purple points, respectively, and depict the range between 
the 16$^{\rm th}$
and 84$^{\rm th}$ percentiles in both \pde{} and \Ytot{} as solid unfilled 
rectangles of the same color. For the reader's convenience,
we tabulate these values in \autoref{t:pdeYtot}.
For comparison, we also show empirical \Ytot{} 
estimates inferred from the literature results of
PHANGS \citep[orange dot-dashed box, ][]{sun2020}, EDGE-CALIFA 
\citep[blue dot-dashed box, ][]{barreraballesteros2021}, and 
KINGFISH \citep[green dot-dashed box, ][]{herreracamus2017}. 
We show the same literature results in each panel, but note that
these measurements are taken on the scale of $\sim 1$ kpc. We also
note that the literature measurements take \HH{} into 
account in their computation of \pde{}.

Directing the reader's attention first to the top and
middle rows (500 pc scales and kiloparsec scales, respectively), we see 
that both the empirically-derived \Ytot{} measured from the UDG sample and 
NSA sample are in good agreement with the predictions of \citeok{}.
We moreover
see that the empirically-derived \Ytot{} of both dwarf samples at
kiloparsec and lower scales
are in good agreement with the measured \Ytot{} values of the 
literature samples of more massive galaxies (which do have \HH{} incorporated in
their \pde{} estimates). This is to say that, as
quantitatively measured by \Ytot{}, the relationship 
between our estimate of ISM weight and star formation 
appears to be relatively constant
between ``extreme'' (low density) dwarfs,
``normal'' dwarfs, and their much more massive counterparts. 
When we consider globally-averaged values, we find that the median
empirical \Ytot{} measured for the NSA dwarfs drops significantly -- 
this may be suggestive of a significant store of \HH{} in these NSA 
dwarfs, which we will discuss further in \autoref{s:discussion:hh}.

Encouragingly, we find that the empirically-measured \Ytot{}
values derived from the dwarf samples on $\leq 1$ kpc
scales are in good agreement with both the theoretical and
numerical predictions for \Ytot{} presented in 
\cite{ostriker2022}. 
However, a unified 
analysis of a sample that includes high \pde{} regions 
(\pde{} $\sim 10^5k_B$ K cm$^{-3}$) would be necessary to expand
quantitatively upon the numerical expectation
of a \pde{} dependence 
beyond the simple literature comparison presented in 
this work.

\begin{deluxetable*}{rl|cccccc}
\tablewidth{0pt}
\tablecaption{16$\rm ^{th}$, \textbf{50$\rm ^{th}$}, and 84$\rm ^{th}$ percentile values of 
\pde{} and \Ytot{} for the UDG and NSA samples at 500 pc, 1 kpc, and global scales.}
\tablehead{
\colhead{} &
\colhead{} &
\colhead{$P_{\rm DE}^{16}$} &
\colhead{$\mathbf{P_{\rm DE}^{50}}$} &
\colhead{$P_{\rm DE}^{84}$} &
\colhead{$\Upsilon_{\rm tot}^{16}$} &
\colhead{$\mathbf{\Upsilon_{\rm tot}^{50}}$} &
\colhead{$\Upsilon_{\rm tot}^{84}$} \\
\colhead{Sample} &
\colhead{Scale} &
\colhead{$\rm \left[ \frac{10^3 K}{k_B\ cm^3} \right]$} &
\colhead{$\rm \left[ \frac{10^3 K}{k_B\ cm^3} \right]$} &
\colhead{$\rm \left[ \frac{10^3 K}{k_B\ cm^3} \right]$} &
\colhead{$\rm \left[ 10^3 \frac{km}{s} \right]$} &
\colhead{$\rm \left[ 10^3 \frac{km}{s} \right]$} &
\colhead{$\rm \left[ 10^3 \frac{km}{s} \right]$} 
\vspace{10pt}
}
\startdata
UDGs& 500 pc &  $1.32_{-0.30}^{+0.18}$ & $3.42_{-0.22}^{+0.28}$ &  $6.60_{-0.22}^{+0.23}$ & $1.01_{-0.18}^{+0.12}$ & $2.13_{-0.14}^{+0.17}$ & $4.02_{-0.26}^{+0.33}$ \\
    & 1 kpc  &  $1.38_{-0.26}^{+0.12}$ & $3.26_{-0.18}^{+0.21}$ &  $6.29_{-0.17}^{+0.20}$ & $1.17_{-0.20}^{+0.13}$ & $2.38_{-0.13}^{+0.11}$ & $4.27_{-0.24}^{+0.28}$ \\
    & global &  $2.46_{-0.33}^{+0.29}$ & $3.69_{-0.31}^{+0.27}$ &  $5.36_{-0.37}^{+0.39}$ & $1.08_{-0.15}^{+0.13}$ & $1.79_{-0.16}^{+0.16}$ & $2.75_{-0.26}^{+0.26}$ \\
\toprule
NSA   & 500 pc &$3.03_{-0.43}^{+0.37}$ & $6.06_{-0.28}^{+0.33}$ & $10.55_{-0.66}^{+0.86}$ & $0.57_{-0.07}^{+0.06}$ & $1.44_{-0.10}^{+0.12}$ & $3.16_{-0.27}^{+0.35}$ \\
dwarfs& 1 kpc  &$2.86_{-0.53}^{+0.45}$ & $6.02_{-0.31}^{+0.34}$ & $10.42_{-0.55}^{+0.76}$ & $0.48_{-0.07}^{+0.06}$ & $1.27_{-0.11}^{+0.11}$ & $2.86_{-0.27}^{+0.30}$ \\
      & global &$4.94_{-0.54}^{+0.52}$ & $7.37_{-0.58}^{+0.58}$ & $10.65_{-0.83}^{+1.26}$ & $0.29_{-0.06}^{+0.07}$ & $0.73_{-0.08}^{+0.10}$ & $1.33_{-0.15}^{+0.18}$ \\
\enddata
\end{deluxetable*}\label{t:pdeYtot}

\section{Discussion}\label{s:discussion}

\subsection{Star Formation Efficiency in UDGs}
In \citetalias{paperone}, we demonstrated that UDGs host relatively
low star formation rate surface densities given their apparent \HI{} 
surface densities. In this work, we have gone on to consider whether
this low efficiency star formation is surprising in the context of 
contemporary theories of galactic star formation. As we have detailed in the
preceding section, we find that the pressure-regulated, feedback-modulated
(PRFM) theory of star formation is remarkably successful at predicting
the low \sigsfr{} and SFE(\HI) of the UDGs. The star formation in these ``extreme'' galaxies,
then, can be well-described within the  same framework of PRFM star
formation as can the ``normal'' (NSA) dwarfs and even more massive galaxies 
-- the difference in their star formation behavior can 
be quantitatively attributed to the shallower gravitational
potential due to the 
characteristically diffuse stellar structure of the UDGs.

The bottom panels of 
\autoref{f:PRFMcomparison} demonstrate that the \oten{} model is highly
successful at reproducing the star formation rate surface densities of the UDGs,
even without considering the surface density contribution of molecular gas. 
Furthermore, \autoref{f:PDEYtot} 
shows that the NSA dwarfs and UDGs lie
on the same relation between \pde{} (dynamical equilibrium pressure) and \sigsfr{} 
as do massive galaxies.
These successes indicate that \HI-rich UDGs,
which are largely in the field, need not be forming stars in an exotic manner, 
as has been suggested for 
globular cluster-rich UDGs in high-density 
environments \citep{danieli2021}.

Beyond a direct comparison of observed and
predicted star formation rate surface densities, 
the PRFM framework allows us
to quantitatively explore the relationship between star formation and
galaxy structure in our samples.
We make observational estimates of \Ytot{} in \autoref{s:results:empiricalYtot}
to compare our results to  both theoretical expectations and 
observational measurements of higher-\pde{} systems. We find that the median value of
\Ytot{} is about $1000-2000\kms{}$  (see \autoref{t:pdeYtot}). This is consistent with
theoretical expectations and similar to previous literature results (though there is significant
variation between literature samples, as shown in \autoref{f:PDEYtot}). Our UDG sample is
characterized by somewhat higher values of \Ytot{} and lower values of \pde{} compared to the 
NSA sample, which is also consistent with the expectation that the momentum injection from
supernovae and FUV heating are more efficient in low density environments due to weaker cooling
and less shielding. However, as we will discuss further in the next section, the \pde{} estimates of
the  NSA sample may be biased low due to the neglected contribution from \HH{}. Thus, the main 
takeaway from the empirical \Ytot{} results should be the agreement between theoretical predictions
and the low-SFE(\HI) UDG sample, rather than the variation between the NSA and UDG samples.
\rrr{The concordance between midplane pressure and star formation rate surface densities in the UDG sample
also implies that \rrrtwo{in the absence of an event that strongly raises \pde{} (e.g. the accretion of
enough high density gas to overcome the effect
of the low stellar mass density on \pde{}),}  
\HI-rich UDGs are unlikely to support \rrrtwo{the} bursty and concentrated star formation 
at $z=0$ 
\rrrtwo{that} has been suggested to form UDGs and/or link them to blue compact dwarfs 
\citep{dicintio2017, sanchezalmeida2018}. 
However, a broader view of the mechanisms that trigger and
sustain concentrated bursts of star formation in low-mass galaxies is needed to further explore the link
between these populations.}

Having established the concordance between the results at hand and both  theoretical \&  observational
results from the literature, we can also examine the implication of the composition of \pde{} on 
predictions of  \sigsfr{}.
Although $\mathcal{W}_{\rm gas}$ is \textit{relatively} more
important for the UDGs (than it is for the NSA dwarfs), we also 
note that $\mathcal{W}_{\rm ext} \geq \mathcal{W}_{\rm gas}$ for all of
the galaxies and length scales probed. This indicates that the external
gravitational potential (from stars and dark matter) plays a significant
role in setting the star formation timescale; that is,
gas surface
density alone is not sufficient to fully predict \sigsfr{}. Indeed, a correlation between
dwarf stellar mass surface density and divergence from the Kennicutt-Schmidt relation has
been demonstrated in samples of nearby galaxies \citep[][see also the top row of \autoref{f:PRFMcomparison}]{delosreyes2019}. 
Altogether, these results paint a picture in which galaxy structure, rather than 
gas availability alone, plays a key role in regulating star formation in low-mass
systems.

\begin{figure*}[htb]
\centering     
\includegraphics[width=\linewidth]{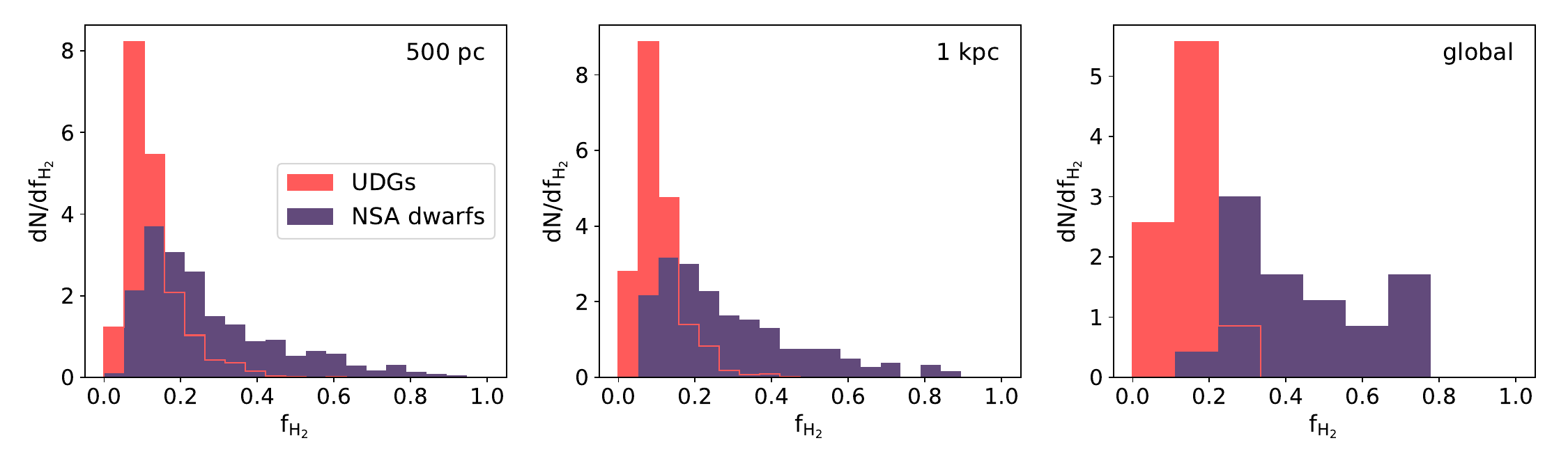}
\caption{ 
    The estimated \HH{} fraction 
    (where f$_{\rm H_2} = \Sigma_{\rm H_2} / [\Sigma_{\rm H_2} +\Sigma_{\rm HI}]$) for 
    the regions in our sample at 500 pc (left), 1 kpc (middle), and global scales (right) when we
    assume a constant \HH{} depletion time of $\tau_{\rm dep} = \Sigma_{\rm H_2}/\Sigma_{\rm SFR} = 1.8$ 
    Gyr. 
    The red histograms shows the UDGs, while the purple histograms show the NSA dwarfs. We find that
    while the distribution in f$_{\rm H_2}$ peaks at f$_{\rm H_2}<0.2$ for both 
    the UDGs and NSA galaxies, the NSA dwarfs are characterized by a tail to high molecular gas fractions.
    This finding is consistent with the under-prediction of the \HI{}-only PRFM estimate for the
    highly star-forming regions of the NSA galaxies (as shown in \autoref{f:PRFMcomparison}).
    }
\label{f:tdep_fh2}
\end{figure*} 

\begin{figure}[htb]
\centering     
\includegraphics[width=\linewidth]{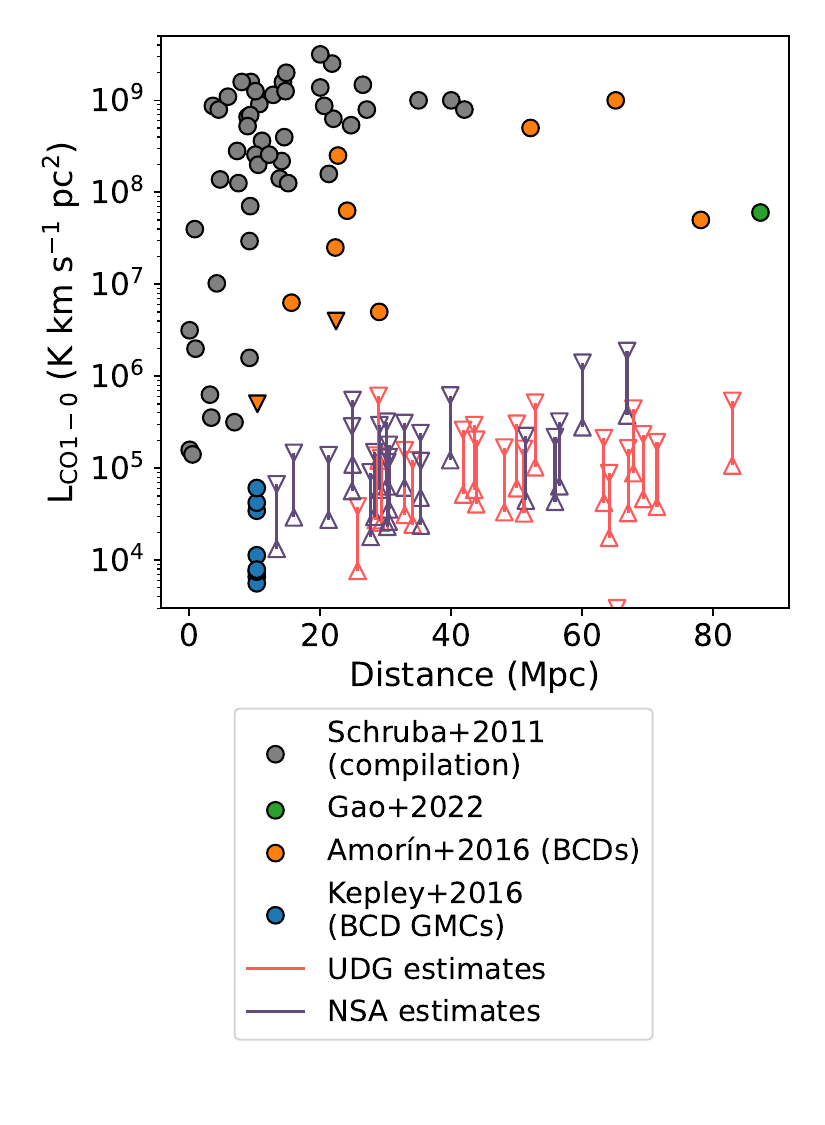}
\caption{ 
    A comparison between the expected CO (1-0) luminosities of our
    NSA (purple) and UDG (red) dwarf samples and real measurements from
    the literature.
    We estimate 
    $\rm L_{CO1-0}$ by assuming a constant \HH{} depletion time of
    $\tau_{\rm dep}=1.8$ Gyr and bounding our estimates from $\alpha_{CO}$
    measurements of the LMC and SMC \citep{schruba2011}. We find that the
    expected CO luminosities of the galaxies in the present sample
    are comparable to measurements made of
    galaxies at lower distances, but lie well below literature measurements of
    galaxies at similar distances.
    }
\label{f:lco}
\end{figure} 


\subsection{ Molecular Hydrogen in UDGs }\label{s:discussion:hh}
In this work we have shown that the PRFM model of star formation does a 
remarkably good job of accounting for the star formation rate surface density of
the UDGs, despite the fact that we assume that the contribution of
molecular hydrogen to the overall weight of the system is negligible. 
In general and quite separate from the discussion of UDGs, 
galaxies in this mass range are thought to be relatively poor in \HH{} overall 
-- a recent study suggests that nearby dwarfs with stellar masses less than 
$\log_{10}(M_\star/M_\odot)=8.5$ have a median \HH{} fraction of 
$f_{\rm H_2} \equiv \Sigma_{\rm H_2} / (\Sigma_{\rm H_2}+\Sigma_{\rm HI}) = 0.15$
\citep{delosreyes2019}. However, studies have also suggested that
dwarfs may be dominated by \HH{} in their central regions, despite being \HH{}-poor
in a spatially-averaged sense \citep{leroy2008}.
Despite their overall small \HH{} content, dwarfs are
also thought to preferentially host their molecular gas at small radii, allowing
\HH{} to contribute significantly to or even dominate the weight of the ISM
near the center of the galaxy \citep{leroy2008}.

In order to understand what we may say about the \HH{} content in UDGs, we 
first consider the impact of our \HI-only
approach on the NSA sample. In \autoref{f:PRFMcomparison}, 
we see that the \HI-only PRFM model significantly
under-predicts \sigsfr{} for the NSA dwarfs, particularly when we consider the
star formation rate surface density averaged over kiloparsec or larger scales.
At 500 pc scales, the bulk of the regions are well-represented by their PRFM
predictions, though a tail towards under-predicted \sigsfr{} at high 
\pde{} remains. This can be understood if one considers that the average \sigsfr{} 
measured within any given region of 500 pc, 1000 pc, or greater size is 
essentially a SFR-weighted average. At global scales, the most vigorously 
star-forming regions dominate the average \sigsfr{}. At 500 pc, these vigorously
star-forming regions are siloed into a relatively small number of points. This 
transition from a roughly SFR-weighted to a roughly area-weighted scheme can also
be seen in the behavior of the median \Ytot{} estimated for the NSA galaxies 
as a function of region spatial scale in the right column of \autoref{f:PDEYtot}.
At 500 pc, the median empirical \Ytot{} computed from the NSA dwarfs is in good 
agreement with the prediction from the TIGRESS simulations of \citeok{}. When 
we consider the global averages, however, the median empirical \Ytot{} lies well 
below the simulation prediction. This behavior is 
consistent with the picture where molecular gas makes
a significant contribution to the weight in the most vigorously star-forming
areas of the NSA dwarfs.

Having established the possible impact on neglecting the
\HH{} contribution in the NSA dwarfs, we may now 
consider the UDGs. First, unlike the NSA dwarfs, the PRFM prediction tends to 
slightly over-predict the star formation rate surface densities of the UDGs as a function of \pde{}. 
This implies that, holding the dark matter halo properties constant (a point
we will return to in the following section), the absence of the weight from the
\HH{} component does not strongly affect the \sigsfr{} predictions for the UDGs. This is despite the fact that, as shown in
\autoref{f:wratio}, the self-gravity of the ISM comprises a
larger component of the total midplane weight of the UDGs. 
The UDGs should thus be relatively more 
sensitive to the effect of neglecting \HH{} gas, as the
$\mathcal{W}_{\rm gas}$ scales as the square of the gas surface
density (as opposed to $\mathcal{W}_{\rm ext}$, which scales
linearly with the gas surface density). Thus, we find that unlike
the NSA dwarfs, the
UDGs in our sample are unlikely to 
host large molecular gas stores.

One simple exercise we can undertake to test the idea that \HH{} is a less
important contributor to \pde{} in UDGs than in the NSA galaxies is by a 
direct estimate of M$_{\rm H_2}$ from the star formation rate via an assumed
depletion time $\tau_{\rm dep}$. Observational support for a constant \HH{}
depletion time of $\tau_{\rm dep} \sim 1 -3$ Gyr has emerged from observational
work
\citep[see, e.g. ][]{bigiel2008,leroy2008, bigiel2011, schruba2011, schruba2012, leroy2021},
though there is evidence for variation within that range as a function of 
environment \citep{utomo2017}. For this test we will assume a constant 
depletion time of $\tau_{\rm dep}\equiv \Sigma_{\rm H_2}/\Sigma_{\rm SFR} = 1.8$ Gyr following the
results of \cite{schruba2011}.

Using this assumed \HH{} depletion time, we may directly estimate the 
expected \HH{} surface density from our measured star formation rate surface densities. We can
then compute the molecular gas fraction in each region,
as shown in \autoref{f:tdep_fh2}, 
where the \HH{} fraction is defined
as f$_{\rm H_2} = \Sigma_{\rm H_2} / (\Sigma_{\rm H_2} +\Sigma_{\rm HI})$. We indeed 
find that the UDG expected \HH{} fractions are quite low, with a median expected molecular gas 
fraction of f$_{\rm H_2}=0.09$ and a 99\thh{} percentile expected fraction of 
f$_{\rm H_2}=0.31$. Conversely, although the median NSA dwarf region has a relatively low
expected \HH{} fraction (f$_{\rm H_2}=0.24$), there is a significant tail to \HH{}-dominated
regions -- the 99\thh{} percentile expected fraction of the NSA sample is f$_{\rm H_2}=0.83$.

Using these \HH{} surface density estimates within the PRFM prediction framework would 
clearly be circular, as the quantity of interest ($\Sigma_{\rm SFR}$) is the very quantity that
we use to estimate $\Sigma_{\rm H_2}$. However, this exercise does provide a separate (though we note,
not entirely independent) consistency check with our previous claim that the molecular gas stores
of the UDGs are a subdominant contributor to the overall gas surface density.

\subsubsection{Outlooks on \HH{} Detection}
Clearly, a direct estimation of the \HH{} content of the UDGs in our sample would
be immensely powerful in determining whether they are \HH-depleted, and whether
their SFE(\HH) is consistent with that of ``normal'' galaxies. There are two 
substantial technical hurdles to gain such an estimation: first, the previously-discussed
low predicted \HH{} masses of the UDGs, and second, the
substantial uncertainty in converting
between a direct measurement (e.g. of CO(1-0)) and \HH{} mass in a 
UDG-like environment. 

Given a depletion time $\tau_{\rm dep}$, the relation between the CO (1-0) luminosity
and \HH{} mass can be written as
\begin{equation}
    \alpha_{\rm CO(1-0)} = \frac{M_{\rm H_2}}{\rm L_{CO(1-0)}} = \tau_{\rm dep}\frac{\rm SFR}{\rm L_{CO(1-0)}}.
\end{equation}\label{e:lco}
As above, we assume an \HH{} depletion time of 1.8 Gyr. 
It is both observed and theoretically expected that \alphaco{} should 
be sensitive to environmental factors such as (but not necessarily limited to) 
metallicity, with \alphaco{} increasing with decreasing metallicity 
\citep[e.g.][]{schruba2011, bolatto2013, gong2020}. Even if a CO detection can be made,
it is therefore also necessary to spectroscopically determine the metallicity 
in order to obtain an estimate of the \HH{} mass, which is in itself observationally
difficult given the low surface brightnesses and on-sky sparsity of the UDGs.
Previous exercises with similar samples have required moderate (1 hr) integration
times with wide (2\arcsec{}) slits
on the 8.1m Gemini South telescope in order to yield gas-phase metallicity 
measurements \citep{greco2018b}.

In the absence of a gas-phase metallicity measurement, we can make a simple estimate of
the expected CO (1-0) luminosity using existing \alphaco{} measurements of our most
nearby massive dwarfs, the Large and Small Magellanic Clouds (LMC and SMC, respectively). 
We adopt the \alphaco{} measurements of the LMC and SMC of \cite{schruba2011} -- as the
more metal-rich system, adopting the \alphaco{} measured for the LMC results in higher
CO (1-0) luminosities than does adopting the value measured from the SMC. 
In \autoref{f:lco}, we show the expected CO (1-0) luminosities of the galaxies in our sample
as a function of luminosity distances given the assumptions detailed above. 
We additionally show a selection of observed CO (1-0) measurements from the literature for 
a selection of samples that include low-mass and low-metallicity objects
\citep{schruba2011, amorin2016, kepley2016, gao2022}. We find that although the 
predicted CO (1-0) luminosities are comparable to $\rm L_{CO1-0}$ measurements of nearby
dwarfs, the predicted luminosities are significantly lower than literature measurements made
at the distance range of the sample. As a small aside, it may be initially surprising to the
reader that the distribution of CO luminosities in the UDG sample is close to that of the 
NSA dwarfs, given that the \HH{} fractions of the UDG sample is much lower. Because we are 
estimating the mass in molecular Hydrogen directly from the integrated star formation rate
(see \autoref{e:lco}), and because the stellar mass range of the samples are chosen to be the
same, this is essentially a restatement of the result that the UDGs lie on the
star-forming main sequence (right panel of \autoref{f:gascomparison}). The lower \HH{} fractions
estimated for the  UDG sample are then a result of elevated \HI{} masses (at fixed stellar mass)
seen in the UDGs versus the NSA dwarfs (left panel of \autoref{f:gascomparison}).

An alternative approach is to measure the dust mass via SED fitting in the 
FIR, where emission may be modeled as a modified blackbody with the dust mass,
temperature, and emissivity index as free parameters 
\citep[see, e.g.][]{kennicutt2011, cigan2021, shivaei2022}. 
Due to the distance and mass of the UDGs, it is likely that
stacking would be necessary to obtain a robust measurement \citep{shivaei2022}.
The dust mass can
then be used to infer to the total gas mass via an assumed dust-to-gas ratio, which
can then in turn be used to estimate the \HH{} mass given a measured \HI{} mass. 
However, like \alphaco{}, the dust-to-gas ratio is naturally expected to depend on 
metallicity; measured dust-to-gas ratios at low metallicities also exhibit non-linearity and
substantial scatter \citep[see Figure 11 of][]{cigan2021}. Thus,
deriving a reliable dust-to-gas ratio for these galaxies
would be challenging even if
metallicity were known.

\subsection{If UDGs Live in Special Halos}
A rather significant uncertainty in our analysis is the unknown 
nature of the dark matter halos in which our \HI-rich UDGs live. A recent work
has shed some light on this topic by estimating halo profile parameters for a 
sample of \HI-rich UDGs \citep{kong2022}. In this work, the authors suggest that
UDGs live in unusually low concentration dark matter halos. Thus, having assumed
``normal'' dark matter halos for the UDGs -- keeping in mind that ``normal'' is
itself an uncertain term for the dwarf population -- we may now consider how
low concentration halos could impact our results.

\begin{figure*}[htb]
\centering     
\includegraphics[width=\linewidth]{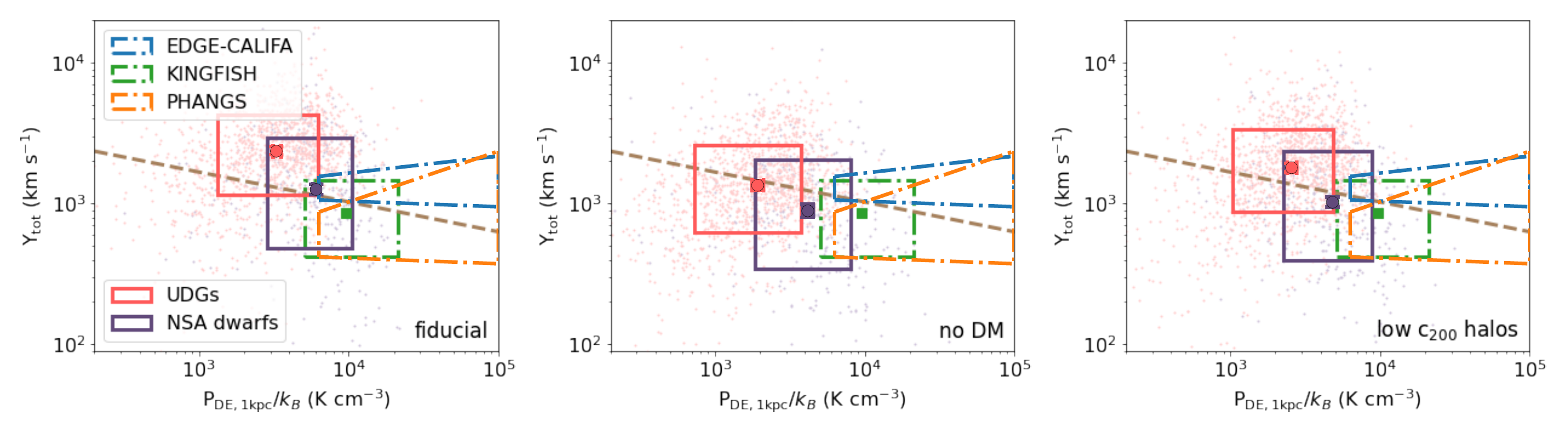}
\caption{
    \rrr{
    The same as the right panel of the middle row of \autoref{f:PDEYtot} with different
    assumptions about the halo properties of our sample. We show our fiducial results at
    left, the empirical feedback yields under the assumption of low concentration halos
    at middle, and the empirical feedback yields with the assumption
    that $\rho_{\rm DM} = 0$ at right. We find that our 1 kpc empirical estimates of the feedback yields under the
    low concentration halo and no-dark-matter assumptions are still consistent with the numerical predictions
    from TIGRESS. Our results are therefore not inconsistent with low dark matter
    fractions or low concentration halos
    in \HI-rich UDGs, though we stress that the precision of our estimates
    precludes us from distinguishing between ``normal halo'' and ``no DM'' modes of
    \HI-rich UDG structure. This exercise also shows that our results are not
    strongly dependent on our fiducial choice of a stellar-to-halo mass relation.
    }
    }
\label{f:nodm_ytot}
\end{figure*}

There is unfortunately only one galaxy, AGC 242019, which is in both our
sample and the sample of \cite{kong2022}. As an exercise, however, we 
estimate the change in the predicted \sigsfr{}  
if we were to adopt the median halo concentration measured by 
\cite{kong2022} of $c=2.62$, which is approximately 30\%
of the median (16\thh{}, 84\thh{} percentile) fiducial halo concentration of $c=9.16\ (8.90, 9.40)$. 
We show the impact of this change in the \pde{}-to-\Ytot{} plane in
\rrr{the middle panel of \autoref{f:nodm_ytot}} -- assuming the lower
concentration halo results in empirical \Ytot{} values
about 25\% lower than the fiducial results. 
We do find that assuming a lower concentration 
halo nominally brings the UDGs closer in line with theoretical results. However, for the
singular galaxy that does constitute the overlap between the samples, we find that
assuming the halo profile reported by \cite{kong2022} actually causes a greater
difference between our observed and predicted \sigsfr{}.
Indeed, the shift in \sigsfr{} predictions as a function
of halo concentration is more emblematic of the 
potential effect of unknown systematics in the \pde{} 
(and empirical \Ytot{}) estimates than it is evidence for
or against unusual dark matter halo profiles in the
UDGs.
There thus yet remains
significant work to be done in order to reduce the uncertainty of the impact of 
the UDGs' dark matter halos.

\rrr{
It has also been suggested that some \HI-rich UDGs are dark matter deficient based on 
observations of their neutral gas kinematics \citep{mancerapina2022}. It is therefore also
useful to understand how the predictions of PRFM theory would change if we were to assume
a dark matter density of $\rho_{\rm DM} = 0.$ For this exercise, we hold all other 
estimates fixed to their fiducial values.}

\rrr{
We find that the midplane pressure estimates of the UDGs are reduced on average to 59\% that of their
fiducial values. The midplane pressure estimates of the NSA galaxies, which we compute for
completeness, are reduced on average to 73\% that of their fiducial values. These reduced values of
\pde{}, and the empirical feedback yields that they imply, remain consistent with PRFM 
predictions at the precision of our estimates, as shown in \autoref{f:nodm_ytot}. We thus do not 
make a statement about whether the $\rho_{\rm DM}=0$ is a better descriptor of UDG star formation,
but rather note that this exercise also indicates that the results presented in this work are not
strongly affected by our fiducial choice of halo parameters. This finding is not
unexpected given that our estimate of the dynamical equilibrium pressure goes 
as the square root of the sum of the stellar and dark matter densities, and that
the self-gravity term is relatively more important for UDGs (see \autoref{f:wratio}).
}

\section{Conclusions}
In this series, we have used spatially-resolved SED fitting to explore
the star formation activity in a sample of nearby (d $<120$ Mpc)
\HI-detected ultra-diffuse dwarf galaxies from \cite{janowiecki2019}, along with
a NASA Sloan Atlas (NSA) reference sample of ``normal'' dwarfs with \HI{} measurements 
from \cite{bradford2015}.
The samples in this work allow us to compare and contrast the star formation behavior of the UDGs
against the NSA dwarfs, providing new tests for star formation theory in extreme (low density) 
environments,
as well as new clues to the evolutionary pathway of these \HI-rich UDGs 
(see \citetalias{paperone}).

As established in the first paper of this series, 
the UDGs are characterized by low star formation rate surface densities and
star formation efficiencies (as a function of their atomic gas surface
densities) down to 500 pc scales. In this work, 
we ask whether the UDGs' lower SFE(\HI) is 
expected in the context of 
contemporary models of star formation. 

We consider the framework of pressure-regulated, feedback-modulated
star formation (\oten{}), which 
directly connects star formation and galaxy structure for 
disks in equilibrium, in \autoref{s:results:sftheory}.
This necessitates an exploration of the dynamical
equilibrium pressure in the UDG and NSA dwarf systems.
The UDGs are characterized by lower stellar $+$ dark matter
densities (\autoref{f:rhosddist}), lower 
dynamical equilibrium pressures (\autoref{f:pdedist}),
and a relatively larger contribution to the overall weight
by self-gravity (\autoref{f:wratio}). 
We indeed find that the lower SFR surface densities and
lower SFE(\HI) seen in the UDGs are well-predicted by the \oten{} model; that is,
the relationship between midplane pressure and star formation rate surface density
(\Ytot) is the same for the UDGs in our sample as it is for the NSA dwarfs, or indeed
even much more massive galaxies. 

This holds true for the UDGs despite the fact that we neglect any weight 
contributions from \HH{} in this analysis. We find that at globally averaged scales 
(which one can think of as the SFR-weighted limit), \HI-only predictions 
underestimate star formation in the NSA dwarfs, but that the same
\HI-only predictions are in good agreement with the NSA dwarfs at 500 pc 
scales (which one may roughly think of as approaching an
area-weighted average). This implies that \HH{} is an important contributor to the
weight in the regions of the most vigorous star formation, but is a minority 
component at large -- a suggestion that is in agreement with previous 
results \citep[see, e.g.][]{leroy2008, delosreyes2019}. 
Star formation in the UDGs, however, is
well-described or even somewhat overestimated by the \HI-only PRFM predictions 
at all spatial scales, suggesting that the UDGs may be \HH-poor even compared to
similarly (stellar) massive ``normal'' dwarfs.

As referenced in \citetalias{paperone}, a clear extension of this work will be
to measure spatially resolved \HI{} in a larger sample of field UDGs. 
In this work, we have also considered
routes to estimate \HH{} mass in the UDGs, 
but note that obtaining a detection via either
CO or FIR dust emission will likely be resource-intensive 
due to the low predicted \HH{} fractions
and likely low metallicities of the UDGs. 
The uncertainty in the $X_{CO}$ factor and/or gas-to-dust
ratio in these galaxies
further increases the uncertainty in obtaining a \HH{} mass from such a detection.
Nevertheless, a greater understanding of the ISM in UDGs is a key path towards
understanding the landscape of star formation in these unusual objects,
as well as towards understanding their utility as laboratories in which to study 
extreme low-density star formation.

\begin{acknowledgements}
\rrrtwo{The authors thank the anonymous referee for their thoughtful, helpful, and thorough 
review of this work.}
The authors thank Eve Ostriker, Song Huang, and Andy Goulding for insightful 
comments and discussion that have greatly improved this manuscript.
The research of 
EKF was supported by the Porter Ogden Jacobus Fellowship. JEG gratefully 
acknowledges support from NSF grant AST-2106730. The work of CGK was supported by NASA ATP grant 80NSSC22K0717.

Based in part on data collected at the Subaru Telescope and retrieved from the HSC data archive system, which is operated by Subaru Telescope and Astronomy Data Center at National Astronomical Observatory of Japan.

The Hyper Suprime-Cam (HSC) collaboration includes the astronomical communities of Japan and Taiwan, and Princeton University. The HSC instrumentation and software were developed by the National Astronomical Observatory of Japan (NAOJ), the Kavli Institute for the Physics and Mathematics of the Universe (Kavli IPMU), the University of Tokyo, the High Energy Accelerator Research Organization (KEK), the Academia Sinica Institute for Astronomy and Astrophysics in Taiwan (ASIAA), and Princeton University. Funding was contributed by the FIRST program from Japanese Cabinet Office, the Ministry of Education, Culture, Sports, Science and Technology (MEXT), the Japan Society for the Promotion of Science (JSPS), Japan Science and Technology Agency (JST), the Toray Science Foundation, NAOJ, Kavli IPMU, KEK, ASIAA, and Princeton University. 

This paper makes use of software developed for the Large Synoptic Survey Telescope. We thank the LSST Project for making their code available as free software at  http://dm.lsst.org

The Pan-STARRS1 Surveys (PS1) have been made possible through contributions of the Institute for Astronomy, the University of Hawaii, the Pan-STARRS Project Office, the Max-Planck Society and its participating institutes, the Max Planck Institute for Astronomy, Heidelberg and the Max Planck Institute for Extraterrestrial Physics, Garching, The Johns Hopkins University, Durham University, the University of Edinburgh, Queen’s University Belfast, the Harvard-Smithsonian Center for Astrophysics, the Las Cumbres Observatory Global Telescope Network Incorporated, the National Central University of Taiwan, the Space Telescope Science Institute, the National Aeronautics and Space Administration under Grant No. NNX08AR22G issued through the Planetary Science Division of the NASA Science Mission Directorate, the National Science Foundation under Grant No. AST-1238877, the University of Maryland, and Eotvos Lorand University (ELTE) and the Los Alamos National Laboratory.

Some of the data presented in this paper were obtained from the Mikulski Archive for Space Telescopes (MAST). STScI is operated by the Association of Universities for Research in Astronomy, Inc., under NASA contract NAS5-26555. Support for MAST for non-HST data is provided by the NASA Office of Space Science via grant NNX13AC07G and by other grants and contracts. 

Based on observations made with the NASA Galaxy Evolution Explorer. GALEX is operated for NASA by the California Institute of Technology under NASA contract NAS5-98034.  

This research has made use of the NASA/IPAC Infrared Science Archive, which is operated by the Jet Propulsion Laboratory, California Institute of Technology, under contract with the National Aeronautics and Space Administration.

This research has made use of the VizieR catalogue access tool, CDS, Strasbourg, France.

\software{Astropy \citep{astropy:2013, astropy:2018}, matplotlib \citep{Hunter:2007}, SciPy \citep{jones_scipy_2001}, the IPython package \citep{PER-GRA:2007}, NumPy \citep{van2011numpy}}, 
pandas \citep{McKinney_2010, McKinney_2011},
Astroquery \citep{astroquery}, extinction \citep{barbary2021}
\end{acknowledgements}

\bibliography{ksdwarfs.bib}

\end{document}